\documentclass[prx,amssymb,amsmath,superscriptaddress,floatfix,twocolumn, nofootinbib]{revtex4-2}
\usepackage{braket}
\usepackage{tikz}
\usepackage{multirow}
\usepackage{booktabs}
\usepackage[export]{adjustbox}
\usepackage{graphicx}
\usepackage{afterpage}
\usepackage[page,toc]{appendix}
\usepackage{amsthm}
\usepackage{dsfont}
\usepackage{hyperref}
\usepackage{physics}
\usepackage{balance}
\hypersetup{
	colorlinks=true,
	linkcolor=blue,
	filecolor=magenta,      
	urlcolor=magenta,
	citecolor=teal
}

\usepackage[mathscr]{euscript}
\usepackage{xcolor}
\usepackage[normalem]{ulem}
\usepackage{soul}
\usepackage[dvipsnames]{xcolor}

\newcommand{\outprod}[1]{\ket{#1}\!\!\bra{#1}}

\DeclareMathOperator{\derdelta}{\partial_{\delta}}

\usepackage{afterpage}
\usepackage{algorithm}
\usepackage{algpseudocode}

\global\long\def\bkt#1{\left(#1\right)}
\global\long\def\eqn#1{\begin{align}#1\end{align}}

\global\long\def\sbkt#1{\left[#1\right]}
\global\long\def\cbkt#1{\left\{#1\right\}}
\global\long\def\abs#1{\left\vert#1\right\vert}

\global\long\def\re{\mathrm{Re}}
\global\long\def\im{\mathrm{Im}}
\global\long\def\dd{\mathrm{d}}

\global\long\def\mr#1{\mathrm{#1}}

\global\long\def\eqn#1{\begin{align}#1\end{align}}
\global\long\def\non{\nonumber}

\usepackage{dcolumn}
\usepackage{bm}
\usepackage{blkarray}
\usepackage{xparse}
\usepackage{mathtools}
\usepackage[shortlabels]{enumitem}
\usepackage{chngcntr}
\usepackage{xcolor}

\begin{document}
	\preprint{APS/123-QED}
	\title{Non-Markovian delay-assisted sensing with waveguide-coupled quantum emitters}
	
	\author{Prajit Dhara}%
    \email[]{prajit.dhara@rtx.com}
	\affiliation{Wyant College of Optical Sciences, The University of Arizona, Tucson, AZ 85721}
    \affiliation{Quantum, Photonics, and Computing Group, RTX BBN Technologies, Cambridge, MA 02138}
    \affiliation{Department of Electrical and Computer Engineering, University of Maryland, College Park, MD 20742}

    \author{Isack Padilla}
    \email[]{iacpad0795@arizona.edu}
	\affiliation{Wyant College of Optical Sciences, The University of Arizona, Tucson, AZ 85721}

    \author{Saikat Guha}
    \email[]{saikat@umd.edu}
    \affiliation{Wyant College of Optical Sciences, The University of Arizona, Tucson, AZ 85721}
    \affiliation{Department of Electrical and Computer Engineering, University of Maryland, College Park, MD 20742}
    
    \author{Annyun Das}
    \email[]{annyun@arizona.edu}
	\affiliation{Wyant College of Optical Sciences, The University of Arizona, Tucson, AZ 85721}
    \affiliation{Department of Physics, The University of Arizona, Tucson, AZ 85721}
	\author{Kanu Sinha}
    \email[]{kanu@arizona.edu}
	\affiliation{Wyant College of Optical Sciences, The University of Arizona, Tucson, AZ 85721}
    \affiliation{Department of Physics, The University of Arizona, Tucson, AZ 85721}    
	\begin{abstract}
        
We show that in a minimal setup of two waveguide-coupled quantum emitters, separated by long distances and subject to an external field, time-delayed feedback can be a resource for sensing field gradients. While the field gradient induces a detuning between the emitters; the large interatomic separations  render the system dynamics non-Markovian. We show that the quantum Fisher  information (QFI) for estimating the detuning parameter, and thereby the field gradient,  is enhanced in the presence of non-Markovian delay. Such an enhancement can be attributed to  the formation of atom-photon quasi-bound states that enable the field to interact with the emitters for longer times, thereby gaining more information about their relative detunings. Additionally, in the presence of delay, the interaction between the emitters is mediated via multiple spectral modes of the field,  further enhancing the sensing capabilities of the system. Our results establish non-Markovian time-delayed feedback and multimode reservoirs  as a resource for distributed quantum sensing with waveguide-coupled quantum emitters.
\end{abstract}
	\maketitle

    \section{Introduction}

\begin{figure}[t]\includegraphics[width=1\linewidth]{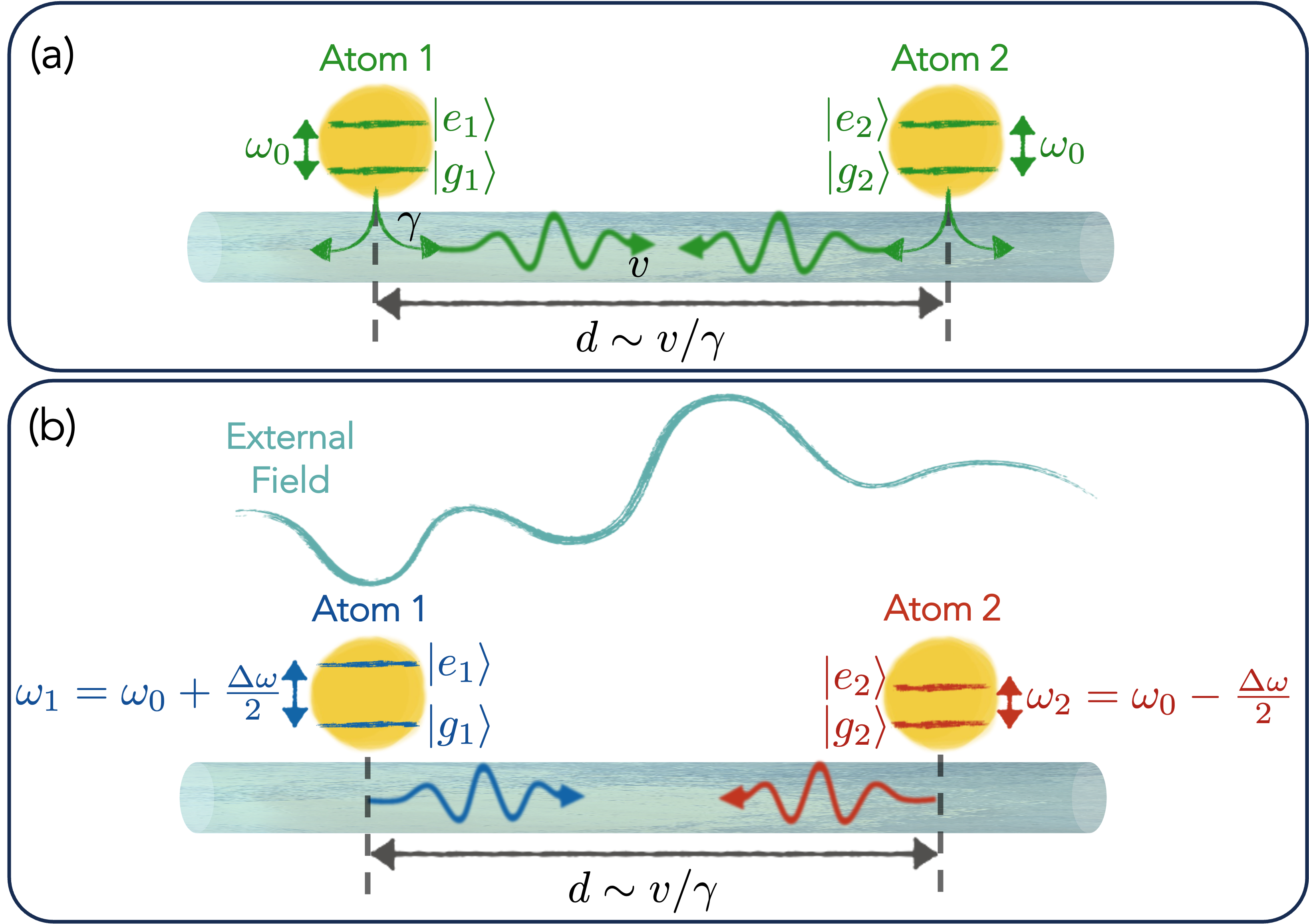}
        \caption{ (a) Two resonant two-level atoms coupled to a waveguide, with their mutual separation $ d$ comparable to the coherence length of a spontaneously emitted photon $ d\sim v/\gamma $, where $ v$  is the speed of the electromagnetic (EM) field in the waveguide, and $ \gamma $ is the spontaneous emission rate of an individual atom into the waveguide.  (b) A spatially-varying external field gradient creates a detuning between the emitters, with resulting resonant frequencies $\omega_1 = \omega_0 + \frac{\Delta \omega}{2}$ and $ \omega_2= \omega_0 - \frac{\Delta \omega}{2}$ for the emitters located at positions $ x_1 = -d/2$ and $ x_2  = d/2$ along the waveguide, respectively.
        }
        \label{fig:sch}
    \end{figure}

Collections of quantum emitters coupled to waveguides are integral to physical implementations of distributed quantum information protocols~\cite{Kimble2008,Sipahigil2016,Javadi2015-te,Faez2014-ek,Laucht2012, AsenjoGarcia2017,Sayrin2015,Konyk2019,Corzo2019}. In such waveguide quantum electrodynamics (QED) systems, as the characteristic delay  between quantum emitters becomes comparable to the coherence length of a spontaneously emitted photon $ (d\sim v/\gamma)$, the memory effects of the field environment predominate the system dynamics (see Fig.~\ref{fig:sch}). Waveguide-coupled emitters in presence of delay exhibit rich non-Markovian collective dynamics with instantaneous collective spontaneous emission rates exceeding those of regular superradiance~\cite{Sinha20a, Dinc2019, Dinc2019exactmarkoviannon}, formation of atom-photon bound states in the continuum (BICs)~\cite{Calajo2019, Facchi2019, KS2019, Carmele2020, Barkemeyer2021, Trivedi2021, Magnifico2025}, and multimode spectral effects~\cite{Sinha20b, Arranz2021, cilluffo2024,  Das2025, Keefe2025}.
In quantum information protocols with significant delays between emitters, time-delayed feedback is not only an essential element of consideration, but it can act as a resource, e.g., for  generation of entangled atomic and photonic  states~\cite{Pichler2017, Pichler2016, Giron, Zheng, capurso2025},  single-photon sources with improved coherence and indistinguishability ~\cite{Crowder2024}, and quantum control via coherent time-delayed feedback~\cite{Grimsmo2015}.

Given that a structured, non-Markovian environment can enhance coherence by enabling information backflow, one can ask~\cite{Breuer2016, deVega2017}: Can the memory effects of a non-Markovian field environment serve as a resource for sensing in waveguide QED systems?
Furthermore, from a metrological perspective, for emitter separations comparable to or larger than the coherence length ($d\gtrsim v/\gamma$), the photonic environment acts as a multimode reservoir with an effective number of modes that scales as $ \sim \gamma d/v$~\cite{Sinha20b, Das2025}. This further motivates the question: Is it possible to leverage the multiple field environment modes as probes to enhance the sensitivity of parameter estimation?

In this work, we show that non-Markovian time-delayed interactions between waveguide-coupled emitters can assist in distributed quantum sensing of field gradients. We consider a minimal setup of two quantum emitters coupled to a one-dimensional waveguide, subject to an external field gradient, which  induces a detuning between the emitters.  We show that in such a system non-Markovian delay is instrumental in enhancing the quantum Fisher information (QFI)~\cite{Helstrom1969} for estimating the field detuning, thereby improving measurement sensitivity. The resulting QFI, in general, is several orders of magnitude higher than that achieved by non-interacting emitters. 

We further show that the enhancement in sensitivity, which occurs for specific values of detunings and delays, coincides with the formation of quasi-bound states of the atoms and photons in the `cavity' formed by the emitters. Such quasi-bound states in the continuum (qBICs) increase the effective memory of the field environment, or equivalently, the number of field modes participating in the collective system dynamics, which translates to the aforementioned sensitivity enhancements. The sensing advantages afforded by our approach are distinct from, and supplementary to the collective enhancements to sensitivity, using an ensemble of quantum systems~\cite{Toth2014-dw,Barry2020-na,Schaffner2024-bu,Muessel2014-li,Li2026-yg,Pezze2018-mz,Kukita2021-lv,Belliardo2026-ni,Brady2026-rz}. Our results provide the first example of environment-assisted sensing in waveguide QED systems, which form the basis for long-distance quantum information protocols.

    The rest of the paper is organized as follows. In Section~\ref{Sec:model}, we present the model of two spatially separated emitters coupled via a waveguide, and subject to an external field gradient that induces a detuning between the atomic resonances. We solve for the atomic dynamics, parametrized by the propagation delay and atomic detuning  in Section~\ref{sec:emitter_dynamics}. In Section~\ref{Sec:fielddyn}, we analyze the  spectrum of the radiated field, identifying the specific parameter values (e.g., propagation delay and detuning) that correspond to an alignment in  In section~\ref{sec:external_sensing}, we study the QFI associated with estimating the relative atomic detuning, identifying specific parameters for which there is an enhancement in sensing. We connect the QFI to the formation of atom-photon quasi-bound states in Section~\ref{sec:bic}, and provide a benchmark case with non-interacting atoms in Section~\ref{sec:bench}. 
    We present our conclusions and outlook in Section~\ref{sec:conc}.

\section{Model}
\label{Sec:model}
    The centerpiece of our analysis is a system comprising a pair of two-level emitters coupled via  a one-dimensional waveguide and interacting with an external field whose gradient we aim to sense (see Fig.~\ref{fig:sch}). 
    The interaction between the external field and the individual emitters shifts their individual resonance frequencies $\omega_0\rightarrow\omega_{1,2}$, such that  $\omega_1-\omega_2=\Delta\omega$. 
    The total Hamiltonian for the joint system may be expressed as $\hat H=\hat H_E + \hat H_F + \hat H_{\rm int}$. Here,
    \begin{align}
        \hat H_E= \sum_{m=1,2}\hbar \omega_m \hat{\sigma}_+^{(m)}\hat{\sigma}_-^{(m)}
        \label{eq:Hamiltonian_emitter}
    \end{align} is the free Hamiltonian for the emitters with  $\omega_m$ as the resonance frequency and $\hat{\sigma}_+^{(m)} = \ket{e_m}\bra{g_m} = (\hat{\sigma}_-^{(m)})^\dagger$ as the atomic raising operator of the $m^{\text{th}}$ emitter. The free Hamiltonian for the waveguide modes is
    \begin{align}
        \hat H_F=  \int_{0}^{\infty} \dd k\,\hbar  \omega\sbkt{\hat{a}^\dagger(k)\hat{a}(k) +  \hat{b}^\dagger(k)\hat{b}(k) },
        \label{eq:Hamiltonian_field}
    \end{align} 
    where $\hat{a}^\dagger(k)$ and $\hat{b}^\dagger(k)$ are the bosonic creation (annihilation) operators for the right-and left-propagating field modes of the waveguide, respectively, and satisfy the canonical commutation relations $[\hat{a}(k),\hat{a}^\dagger(k')]=\delta(k-k')$, $[\hat{b}(k),\hat{b}^\dagger(k')]=\delta(k - k')$. We assume the waveguide to have a linear dispersion such that $k = \omega/v $. The interaction between the emitters and the waveguide modes is described by $ \hat H_{\rm int}$, which, upon making the electric-dipole and rotating-wave approximations, can be expressed in the interaction picture as
    \begin{widetext}
    \eqn{
            \hat H_{\mathrm{int}}=\hbar\sum_{m=1,2}  \int_0^{\infty} \dd\omega\, g(\omega) \hat{\sigma}_+^{(m)}&\sbkt{\hat{a}(\omega)e^{i\omega x_m/v}+\hat{b}(\omega)e^{-i\omega x_m/v}} e^{-i(\omega-\omega_m)t} + \mathrm{H.c.} . 
            \label{eq:Hamiltonian_interaction}
     }        
     \end{widetext}
    Here, $g(\omega)$ is the atom-field coupling strength; we assume a flat spectral density around the resonant frequencies of the emitters such that $g(\omega)\approx g(\omega_0)$. The two emitters are located at $x_1=-d/2$ and $x_2=d/2$ along the waveguide axis, such that the interatomic separation $ d $ can be large and comparable to the coherence length $ \sim v/\gamma$ associated with a spontaneously emitted photon. 
    
    We assume that the initial state of the total system is $\ket{e_1,g_2}\ket{\cbkt{0}}$, wherein emitter 1 is prepared in an excited state while emitter 2 is ground; $\ket{\{0\}}$ denotes the vacuum state for the waveguide modes.  We remark that the system is initialized in an unentangled state, wherein we are not leveraging entanglement as a metrological resource to start with~\cite{Giovannetti2006}. Since the total Hamiltonian preserves the total number of excitations in the atomic + field Hilbert spaces, the system remains in a single-excitation manifold at all times. 
    The joint emitters-field state at any time $t\geq0$ is thus:  
    \eqn{\label{eq:single_excitation_ansatz}
       &\ket{\Psi_{\eta,\delta}(t)} =\sum_{m={1,2}} c_m(t)\hat{\sigma}_+^{(m)}\ket{g_1,g_2,\{0\}}\non\\   &+\int_0^\infty \dd\omega \, \sbkt{c_a(\omega,t)\hat{a}^\dagger(\omega)+c_b(\omega,t)b^\dagger(\omega )} \ket{g_1,g_2,\{0\}},
    }
    where $c_m$(t) represents the excitation amplitude for the $m^\mr{th}$ emitter,  and $c_{a}(\omega,t)$, $c_{b}(\omega,t)$ are the field excitation amplitudes for the right-and left-going waveguide modes, respectively. Here we have parametrized the quantum state of the total system with dimensionless distance  $\eta \equiv \gamma d/v$ and detuning $ \delta \equiv \Delta \omega/\gamma$.  We now consider the dynamics of the total detuned emitters + field system under the Hamiltonian Eq.~\eqref{eq:Hamiltonian_interaction} in the single-excitation subspace. 
    

    \section{Collective Dynamics of Detuned Emitters }
    \label{sec:emitter_dynamics}
    
    We  first trace over the field modes, including the self-consistent backaction of the field on the emitters, to determine the coupled dynamics of the two detuned atoms as follows (see Appendix~\ref{app:emitter_dyn} for details):
    
    \label{eq:atom_dynamics}
            \begin{align}
            \label{eq:dc1dt}
        \dot{c}_1(t)=&-\frac{\gamma}{2}\left[c_1(t) \right.\non\\
        & \left.+ \beta c_2\left(t-\frac{d}{v}\right)\Theta\left(t-\frac{d}{v}\right)e^{i\omega_2 d/v + i\Delta\omega t} \right], \\
        \label{eq:dc2dt}
        \dot{c}_2(t)=&-\frac{\gamma}{2}\left[c_2(t) \right.\non\\
        & \left.+ \beta c_1\left(t-\frac{d}{v}\right)\Theta\left(t-\frac{d}{v}\right)e^{i\omega_1 d/v - i\Delta\omega t} \right],
    \end{align}
    where $\gamma=2\pi|g(\omega_0)|^2$, and $\beta$ represents the coupling efficiency of the emitters to the waveguide, which we will assume to be  $\beta=1$, unless stated otherwise. 
    The first term in the above equations corresponds to  spontaneous decay of each emitter; the second term represents  the waveguide-mediated time-delayed feedback of one emitter on the other while accounting for the appropriate propagation phase $(e^{i \omega_m d/v})$ and the detuning between the emitters $ (e^{i \Delta \omega t})$.
     The  solution to the coupled atomic dynamics in Eqs.~\eqref{eq:dc1dt} and \eqref{eq:dc2dt}  involves inversion of the  Laplace-transformed coefficients, $\mathscr{C}_m(s) =\int_{0}^{\infty} \dd t\,  c_m(t)e^{-st}$:
    \begin{align}
    \label{eq:laplacec1}
				    \gamma	\mathscr{C}_1(s)&=\frac{(\tilde{s}-i\delta+1/2)}{(\tilde{s}+1/2)(\tilde{s}-i\delta+1/2)-\beta^2 ( e^{i\phi_1}e^{-\eta\tilde{s}}/2)^2}, \\
                    \label{eq:laplacec2}
				\gamma	\mathscr{C}_2(s)&=\frac{-\beta  e^{i\phi_2} e^{-\eta\tilde{s}}/2}{(\tilde{s}+1/2)(\tilde{s}+i\delta+1/2)-\beta^2 ( e^{i\phi_2} e^{-\eta\tilde{s}}/2)^2},
    \end{align}
    where we use the normalized variable, $\tilde{s}\equiv s/\gamma$ and the propagation phase associated with each frequency is $\phi_m=\omega_md/v$.  We note that the propagation phases $\phi_{\{1,2\}}$ must satisfy the self-consistent relation
     \begin{align}
         \phi_1-\phi_2= \frac{(\omega_1-\omega_2)d}{v} = \delta\eta.
         \label{eq:phase_self_cons}
     \end{align}
    
     We now analyze the emitter dynamics resulting from the inverse Laplace transform of Eqs.~\eqref{eq:laplacec1} and \eqref{eq:laplacec2} that can be expressed in two different forms: (1) a series solution~\cite{MilonniKnight74, Milonni1975} which illustrates the time-delayed feedback on the atomic dynamics from the field environment; or (2)  a multi-exponential decay~\cite{Sinha20a} which illustrates the multimode nature of the field environment in presence of retardation, as we will discuss further below. 
    
     \subsection{Series solution}

The atomic dynamics can be written in terms of a series solution as follows (see Appendix~\ref{appendix:series_solution} for details):
     \begin{widetext}
         
     \begin{align}
     \label{eq:c1t}
             c_1(t)
    =&   e^{-\gamma t/2} +   e^{i\delta \gamma t/2}\sum_{n=1}^\infty \Xi_n^{(1) } (t)\beta^{2n} e^{in(\phi_1+ \phi_2)}e^{-(\gamma t-2n\eta)/2} \Theta(\gamma t-2n\eta),\\
    \label{eq:c2t}
    c_2(t) 
    =& e^{- i \delta \gamma t/2} \sum_{n=0}^\infty\Xi_n^{(2)}(t)  \beta^{2n+1} e^{i(2n+1)(\phi_1 + \phi_2)/2}  e^{ - \bkt{\gamma t  - \bkt{2n + 1}\eta}/2} \Theta[\gamma t-(2n+1)\eta] .
\end{align}
     \end{widetext}
 Each term in the above series corresponds to the modification of atomic dynamics after a round trip of the field between the emitters, as seen from the Heaviside theta functions $\cbkt{\Theta \bkt{\gamma t - 2 n \eta} ,\, \Theta \bkt{\gamma t - (2 n + 1)\eta} }$, which make the non-Markovian  time-delayed feedback manifest. The first term in the atom 1 dynamics $ c_1 (t)$ corresponds to its independent decay and the second term to the time-delayed feedback from atom 2 at  every $t = 2nd/v$ that incorporates the waveguide coupling efficiency $ (\sim \beta ^{2n})$ and the round-trip propagation phase $ e^{i n \bkt{\phi_1 + \phi_2}}$. Similarly, the excitation dynamics of atom 2 results exclusively due to the time delayed-feedback from atom 1 at every $t = (2n+1)d/v$. The dynamics of atom 1 (2) as `driven' by atom 2 (1), exhibits oscillations with frequencies $ \Delta \omega/2  =  \delta \gamma/2$ $(-\Delta \omega/2  =  -\delta \gamma/2)$~\cite{Lee2023}. The terms $ \Xi _n ^{(m)}(t)$ correspond to the non-exponential part of the atomic dynamics, their explicit expressions are given in Appendix~\ref{appendix:series_solution}.
    
\begin{figure}[t]
        \centering
        \includegraphics[width=\linewidth]{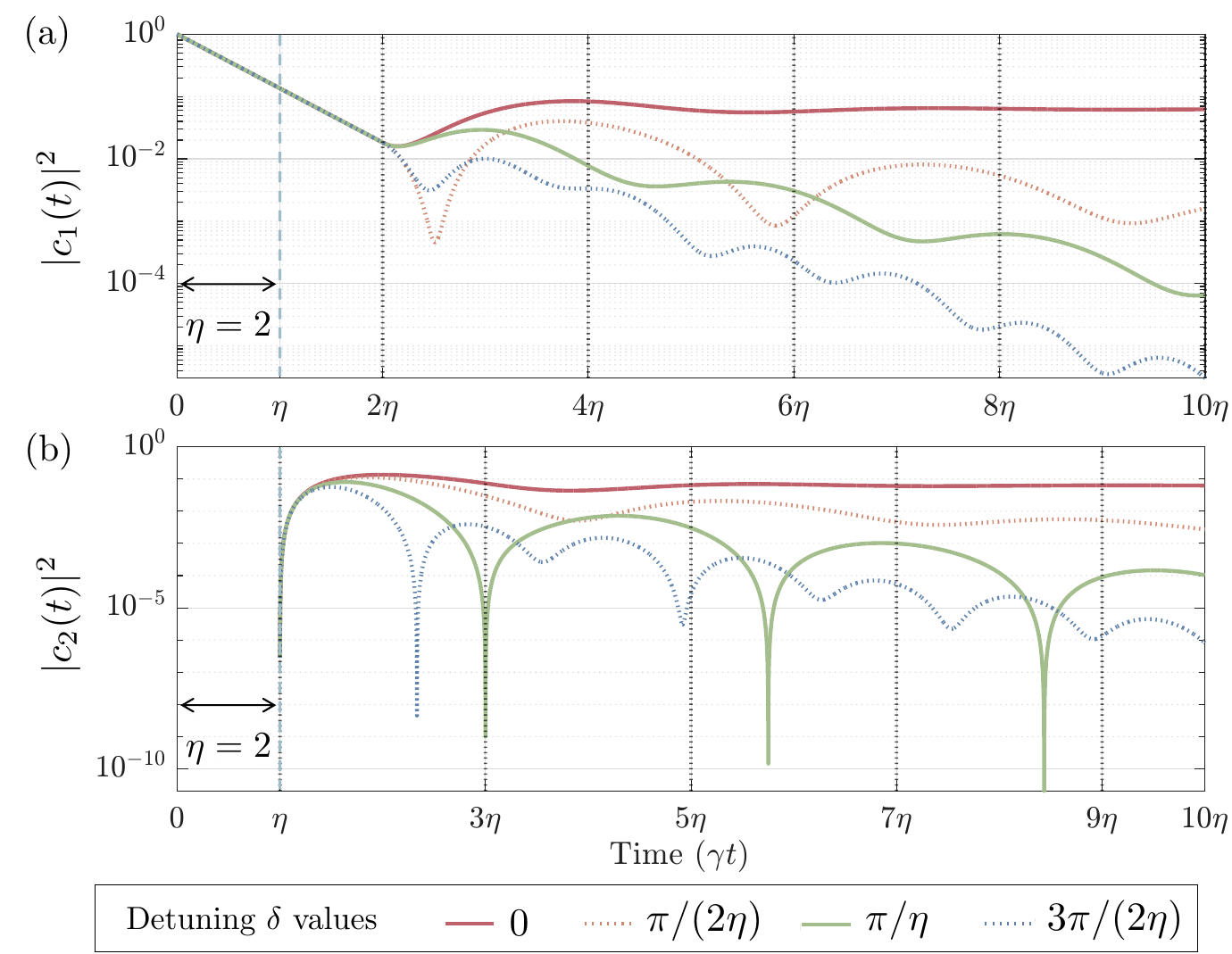}
        \caption{
        Emitter and field dynamics in the presence of detuning  for the initial state $\ket{e}_1\ket{g}_2\ket{\cbkt{0}}$. Emitter excitation probabilities (log-scaled) for emitters 1 (a) and 2 (b) as a function of time ($\gamma t$)  and interatomic separation $\eta = 2$ for detuning values $ \delta = \cbkt{0, \frac{\pi}{2\eta }, \frac{\pi}{\eta }, \frac{3\pi}{2\eta }}$. The vertical gray dotted lines represent the time intervals for the field excitation round-trip between the two emitters ($ t = n d/v$ or $ \gamma t = n \eta$).
        }
        \label{fig:emitter_dynamics}
    \end{figure}
   
    \subsection{Multiexponential decay}
    We invert $\mathscr{C}_i(s)$ based on Cauchy's residue theorem to obtain the atomic dynamics expressed as a multiexponential decay (see Appendix~\ref{App:multiexp} for details):
    \begin{align}
        c_m(t) 
        &\equiv \sum_{\boldsymbol{j}}{R}_{\boldsymbol{j}}^{(m)} e^{\tilde{s}_{\boldsymbol{j}}^{(m)} \gamma{t}},
        \label{eq:pole_solution}
    \end{align}
    where  $\tilde{s}_{\boldsymbol{j}}^{(m)}$ and $R_{\boldsymbol{j}}^{(m)}$ are the poles and residues corresponding to the inverse Laplace transform of $\mathscr{C}_m(s)$, evaluated semi-analytically. 
    For practical purposes, one may truncate the pole index summation to a finite cutoff value $|j|\leq j_{\rm cut}$; higher order poles capture larger frequencies with  progressively smaller contributions~\cite{Sinha20a}.


The multiexponential decay form of the emitter dynamics alludes to a multimode field environment that mediates the interatomic interactions. Physically, this can be understood as follows: the `cavity' formed by the two emitters has a free spectral range  \eqn{ \Delta_\mr{FSR}\equiv \frac{\pi v}{d};} thus as $ \eta \sim \gamma/\Delta_\mr{FSR} \gtrsim 1$,  the atomic linewidth encompasses multiple such cavity modes~\cite{Das2025, cilluffo2024, Keefe2025}.  
As we will show, such  multimode dynamics, mathematically equivalent to the non-Markovian time-delayed feedback, is crucial to enhancing the detuning sensitivity highlighted in Sec.~\ref{sec:external_sensing}. In essence, we demonstrate that increasing the number of environment degrees of freedom participating in the system dynamics, or equivalently extending the environment's memory time, can enhance the sensitivity for estimating system parameters.
    
\subsection{Detuned emitter  dynamics}

The introduction of detuning modifies the collective emitter dynamics, as noted previously in Eqs.~\eqref{eq:c1t} and~\eqref{eq:c2t}. We highlight this in Fig.~\ref{fig:emitter_dynamics}, where we plot excitation probability dynamics ($|c_1(t)|^2$, $|c_2(t)|^2$) for the two emitters as function of time $(\gamma t)$ with a fixed separation $ \eta = 2$, and different values of the detuning $ \delta$ chosen as $\delta=n\pi/(2\eta)$  ($n\in \mathbb{Z}^+$). We choose the propagation phase  \eqn{\label{eq:phi2}\phi_2 =   2 p \pi ;\quad p \in \mathbb{Z^+},} for the resonantly emitted field from emitter 2, such that $ \omega_2 = 2 p \Delta_\mr{FSR}$, which coincides with one of the geometric resonances of the cavity formed by the emitters.  The corresponding propagation phase for the resonantly emitted field from atom 1 becomes:
\eqn{ \phi_1  = 2 p \pi + \delta \eta 
\implies\omega _1 = \left(2p + \frac{n}{2}\right) \Delta_\mr{FSR}.}
Thus, for even values of $n$  we have the resonance of atom 1  coincide with one of the `atomic cavity' resonances; while for odd values of $n$, there is  an offset between the resonance of emitter 1 and that of the atomic cavity. 
    \begin{figure*}
        \centering
        \includegraphics[width=1 \linewidth]{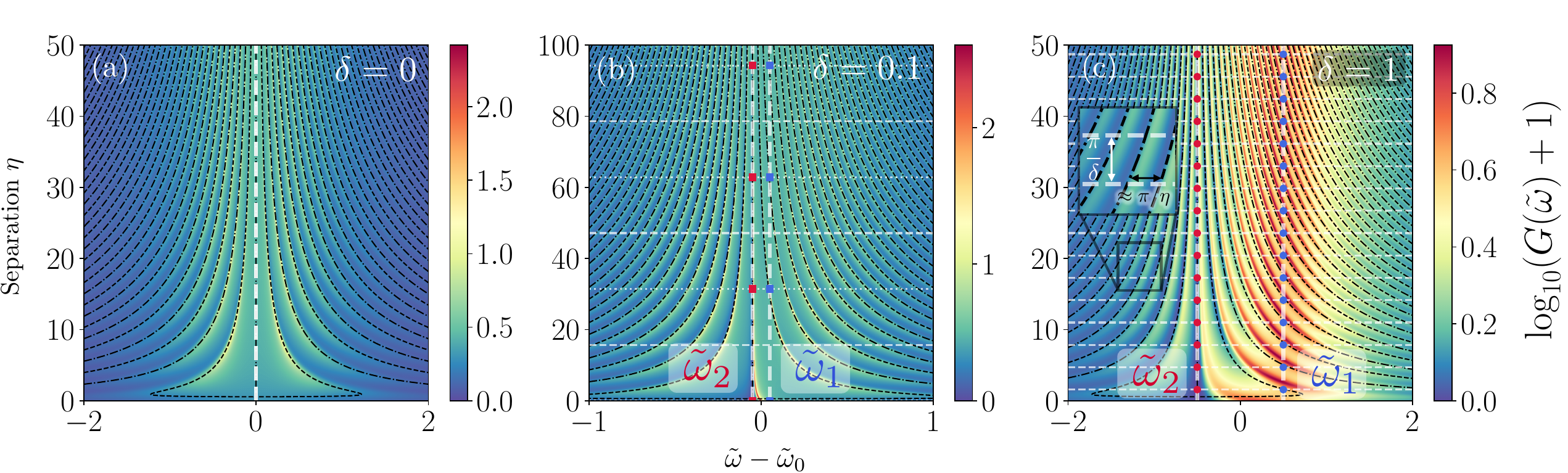}
        \caption{Field spectrum $G(\tilde \omega)$ as a function of the frequency $ \tilde \omega- \tilde \omega _0$ and the interatomic separation $ \eta$ for different values of detuning: (a) $\delta=0$,  (b) $\delta=0.1$, and (c) $\delta=1$.  The vertical dashed lines marked as $ \tilde \omega_{1,2}$ indicate the resonances of the two emitters. The peaks of the spectrum obtained from Eq.~\eqref{eq:baromega_pm_n} are denoted by the black dash-dotted ($ \tilde \omega_{n\delta} ^+ - \tilde \omega_0 $) and black dashed ($ \tilde \omega_{n\delta} ^- - \tilde \omega_0 $) curves. The horizontal white dotted and dashed lines in (b) and (c) represent specific values of $\eta_n = n \pi/\delta$ and $ \eta_{n + 1/2} = \bkt{n + 1/2}\pi/\delta$, respectively. The blue (red) squares and circles indicate the intersections of the horizontal $\eta = \eta_n $ and $\eta = \eta_{n + 1/2} $ lines with the atomic resonances $\tilde \omega_1 - \tilde \omega_0 $ ($\tilde \omega_2 - \tilde \omega_0 $), respectively. The highlighted inset region in (b) shows the separation between the adjacent horizontal white $ \eta_n $ lines is $ \pi /\delta $ for a fixed $ \tilde \omega$ (Eq.~\eqref{eq:deltaeta}); while for a fixed (large)  separation $ \eta$, the free spectral range  between two adjacent spectral peaks is $ \approx\pi/\eta $ (Eq.~\eqref{eq:spectrumFSR}), akin to a Fabry-P\'erot cavity. }
        \label{fig:spectrum}
    \end{figure*}
 Fig.~\ref{fig:emitter_dynamics}(a) and (b) show that the dynamics of emitter 1(2) is influenced by the time-delayed feedback  from emitter 2(1) every $\gamma t$ that is an even (odd) multiple of $\eta$, in agreement with Eqs.~\eqref{eq:c1t} and \eqref{eq:c2t} (time intervals marked by gray vertical lines).  For resonant emitters ($\delta = 0$, red solid line), the emitters have a finite excitation probability in the steady state. This can be understood as follows: the initial state of the emitters $ \ket{e_1, g_2}$ is an equal superposition of a superradiant ($ \frac{1}{\sqrt2}(\ket{e_1, g_2 }+ \ket{g_1, e_2})$) and subradiant ($ \frac{1}{\sqrt2}(\ket{e_1, g_2 }- \ket{g_1, e_2})$) state; while the symmetric superradiant part radiates into the field modes, the anti-symmetric dark state forms an atom-photon bound state~\cite{Calajo2019, Sinha20a, KS2019}. 
 Detuning between emitters prevents the formation of a bound steady state. As a notable feature, the rate of change in atom 1 excitation probability as it `sees' a retarded backaction from atom 2, determined by the slope of $|c_1(t)|^2$ at $\gamma t=2\eta$, is positive (negative) when the detuning  $ \delta$ is an even (odd) multiple of $\pi/(2\eta)$. Thus, for a fixed separation, the choice of detuning determines the revival of atomic excitation, and concomitantly the emission from the atoms into the field. We will see that  detunings, $ \delta $, that are odd integer multiples of $\pi/(2\eta)$, lead to the formation of a quasi-bound state  of emitters and the field in the `atomic cavity' region.  In subsequent section, we examine the field spectrum, which supports this observation and, we will show later in Section~\ref{sec:external_sensing}, assists in field gradient sensing.

\section{Field Spectrum}
~\label{Sec:fielddyn}

    We now analyze the spectrum of the field radiated by the atoms. Given the non-Markovian nature of the dynamics the field environment plays a critical role in the  collective dynamics of the atoms, and consequently in sensing the relative atomic detuning.  

    The field excitation amplitudes in the steady state ($t\to\infty$)  can be written as (see Appendix~\ref{app:emitter_dyn}):
    \begin{align}
        c_a(\tilde\omega,\infty)&=-i\sqrt{\frac{\gamma}{2\pi}} \left[e^{i\tilde\omega \eta/2} F_1(\tilde\omega) + e^{-i\tilde\omega \eta/2}  F_2(\tilde\omega)\right],
        \label{eq:ca_inf}\\
        c_b(\tilde\omega,\infty)&=-i \sqrt{\frac{\gamma}{2\pi}} \left[e^{-i\tilde\omega \eta/2} F_1(\tilde\omega) + e^{i\tilde\omega \eta/2}  F_2(\tilde\omega)\right],
        \label{eq:cb_inf}
    \end{align}
    where $\tilde\omega=\omega/\gamma$. The atomic frequency response functions,
    $F_1(\tilde\omega)$ and $F_2(\tilde\omega)$, are defined as the Fourier transforms of the atomic coefficients $ F_{m}(\tilde \omega)=\int_0^\infty \dd  t\, c_m(t) e^{i( \tilde\omega -\tilde \omega_m )\gamma t}$, which describe the frequencies present in the atomic dynamics.     We  note from Eqs.~\eqref{eq:ca_inf} and \eqref{eq:cb_inf} that the field excitation amplitudes $( c_{a,b}(\tilde \omega, \infty))$ result from the interference of the fields radiated by the two emitters, incorporating the propagation phase $(e^{\pm i\tilde\omega\eta/2})$ between the emitters. The response functions $F_m\bkt{\tilde \omega}$  are given by (see Appendix~\ref{appendix:atomic_frequency_response} for details):
    \begin{align}
        \gamma F_1(\tilde\omega) &= \frac{-i[(\tilde{\omega}-\tilde\omega_0)+\delta/2]+1/2}{D(\tilde\omega)},\label{eq:f1} \\
        \gamma F_2(\tilde\omega) &= -\frac{ e^{i\eta \tilde\omega}/2}{D(\tilde\omega)},
        \label{eq:f2}
    \end{align}         
with the denominator $D(\tilde \omega)$:
\begin{align}
    D(\tilde\omega)=\sbkt{-i(\tilde\omega-\tilde\omega_0) + \frac{1} {2}}^2+\frac{\delta^2}{4} - \frac{e^{2i\eta \tilde\omega}}{4}.
    \label{eq:dw}
    \end{align}
The roots of $D(\tilde \omega)$ capture the characteristic frequencies present in the dynamics of the emitters. Appendix~\ref{appendix:atomic_frequency_response} shows that the peak frequencies in the dynamics of atom 1 are centered at $ \tilde \omega_1$, and those for atom 2 are located symmetrically at $ \tilde\omega_1$ and $ \tilde\omega_2$ (see Fig.~\ref{Fig:F1F2}).

    The steady-state spectrum  of the emitted field can be defined as
    \eqn{G\bkt{\tilde \omega}\equiv  |c_{a}(\tilde \omega,\infty)|^2 + |c_{b}(\tilde \omega,\infty)|^2,
    \label{eq:gw}
    }
    which represents  the probability of excitation of a  field mode with frequency $\tilde \omega$. 
    This can be expressed in terms of the atomic frequency responses as:

    \eqn{
    G\bkt{\tilde \omega} =  & \frac{\gamma}{\pi} \sbkt{\abs{F_1 (\tilde\omega)}^2 + \abs{F_2 (\tilde\omega)}^2\right.\non\\
    &\left.+ 2 \cos \bkt{\tilde \omega\eta}\re\sbkt{F_1 ^\ast \bkt{\tilde \omega}F_2 \bkt{\tilde \omega}} }.
    \label{eq:Gbarw}
    }



    Fig.~\ref{fig:spectrum} shows the emitted field spectrum $ G\bkt{\tilde \omega }$ as a function of frequency $\tilde\omega$ and the interatomic separation $\eta$, for three separate values of the detuning $ \delta$. 
    We highlight the salient features of the field spectrum as follows:
    \begin{itemize}
        \item {For resonant emitters $(\delta=0)$, the spectrum peaks are obtained via the zeros of the denominator of Eqs.~\eqref{eq:f1bar}-\eqref{eq:f2bar} ($D(\tilde\omega_{n0}^\pm)=0$) as follows~\cite{Sinha20b}:      
    \eqn{
        \tilde \omega_{n0}^\pm = \tilde\omega_0+\frac{1}{\eta}{\rm Im}\left[W_{n}\left(\pm\frac{\eta}{2}e^{\eta/2}\right)\right],\,\, n\in\mathbb{Z} ,
        \label{eq:omega_pm_n}
    }
     where $W_{n}(\cdot)$ is the $n^{\text{th}}$ branch of the Lambert-W function. We have assumed a propagation phase of $ \phi_0 = \phi_1 = \phi_2 = 2 p \pi$ ($p \in \mathbb{Z}^+$) for resonant emitters. Fig.~\ref{fig:spectrum}\,(a) shows the spectrum of the field emitted from two resonant emitters, where we note that the peaks of the emission coincide with the black dash-dotted and dashed curves representing $ \tilde \omega_{n0} ^+ - \tilde \omega_0 $ and $ \tilde \omega_{n0} ^- - \tilde \omega_0 $, respectively.
}
    \item {For detuned emitters, we observe from Fig.~\ref{fig:spectrum}\,(b) ($\delta = 1$) and Fig.~\ref{fig:spectrum}\,(c) ($\delta = 2$) that, as expected, the field emission from initially excited atom 1  is peaked at its resonance frequency $ \tilde \omega_1$. For large $\eta$, the peaks of the emission spectrum can be approximated by:
    \eqn{
    \tilde \omega_{n\delta} ^\pm =  \tilde\omega_0 - \frac{\delta }{2}  +\frac{1}{\eta}{\rm Im}\sbkt{W_{n}\left(\pm\frac{\eta}{2}e^{\eta/2}\right)},\,\, n\in\mathbb{Z} ,
        \label{eq:baromega_pm_n}
    }
    as seen from the black dash-dotted and dashed curves in  Fig.~\ref{fig:spectrum}\,(b) and (c). The above expression reduces to Eq.~\eqref{eq:omega_pm_n} for $ \delta = 0$.}
    \item{When the emitter 1 resonance $ (\tilde \omega_1)$ is coincident with the  resonances of the atomic `cavity', such that   $\tilde\omega_1=\tilde \omega_{n\delta}^\pm$, we observe a stronger overall emission, as indicated by the blue squares and circles in Fig.~\ref{fig:spectrum}(b) and (c), respectively. This occurs for specific values of the atomic separation $ \cbkt{\eta_{n + 1/2},\eta_n } $, depending on the relative atomic detuning $ \delta $ (see Appendix~\ref{appendix:eta_delta_condition} for details):

    \eqn{\label{eq:deltaeta}
    \eta_{n + 1/2} \delta &\approx \bkt{n + \frac{1}{2}} \pi, \quad \text{for $ \delta\gtrsim 1$};\\
    \label{eq:deltaetasmall}
    \eta_n \delta &\approx n \pi, \quad \text{for $ \delta\ll 1$}.    
    }

    \item The difference between the subsequent resonances of the emitted field spectrum, $\abs{\tilde\omega_{n\delta}^+-\tilde\omega_{n\delta}^- }$, converges to   (see Appendix~\ref{appendix:eta_delta_condition}):
    \eqn{
    \abs{\tilde\omega_{n\delta}^+-\tilde\omega_{n\delta}^- }\rightarrow\frac{\pi}{\eta} = \frac{\Delta_\mr{FSR}}{\gamma}
    \label{eq:spectrumFSR}}
    in the large $\eta$ limit, recovering the FSR of a Fabry-P\'erot cavity of length $d$~\cite{MeystreBook2007}.
}
    \end{itemize}
We will see in the next section that the conditions $\eta_{n+1/2} \delta\approx (n+1/2)\pi$ (for $ \delta \gtrsim 1$) and $ \eta _n \delta \approx n \pi $ (for $ \delta \ll1$), where there is an alignment of the atom 1 resonance with that of the `cavity' region between the atoms, there is a formation of atom-photon quasi bound states.  Thus, the excitation is kept for longer times in the form of field build up inside  the two-atom `cavity'. We hypothesize that such quasi bound states, which are quintessentially non-Markovian,  can potentially enhance sensing, as we will now demonstrate.
 
    \section{Sensing External Fields}
    \label{sec:external_sensing}
            Having developed a framework to evaluate the quantum state of the emitters and the field environment, we now evaluate the  QFI  associated with sensing the relative atomic detuning  $\delta$. The QFI bounds the highest precision in estimating $\delta$ attainable by any measurement as per the quantum Cram\'er-Rao bound~\cite{Braunstein1994}. 
    \subsection{Quantum Fisher Information}
 
    Throughout our analysis, we assume perfect emitter-waveguide coupling ($\beta=1$), which implies that the state of the total atom + field system remains pure at a general time $ t>0$: \eqn{\rho_{\delta,\eta}(t) \equiv \outprod{\Psi_{\delta,\eta}(t)}.}
    For a pure state, the QFI, $H(\delta)$, can  be simplified as~\cite{Fujiwara1995-kz}:
    \begin{align}
       \begin{split}
           &H(\delta; t) =  4 \Tr \rho_{\delta,\eta}(t) \left(\frac{\partial \rho_{\delta,\eta}(t)}{\partial \delta}\right)^2\\
            = & 4 \left[ \mathrm{Re} \Braket{\partial_\delta \Psi_{\delta,\eta}(t)|\partial_\delta \Psi_{\delta,\eta}(t)}+ \Braket{\Psi_{\delta,\eta}(t)|\partial_\delta {\Psi_{\delta,\eta}(t)}}^2  \right],
            \label{eq:qfi_definition}
       \end{split}
    \end{align}
    where $\ket{\partial_{\delta}\Psi_{\delta,\eta}(t)}$ is the covariant derivative of $\ket{\Psi_{\delta,\eta}(t)}$ with respect to $\delta$ on the manifold of pure states\footnote{It is important to note that $\ket{\partial_{\delta}\Psi_{\delta,\eta}(t)}$ and $\bra{\partial_{\delta}\Psi_{\delta,\eta}(t)}$ are differential unit vectors defined for the state $\ket{\Psi(t)}$ and its dual; they are not necessarily self dual, i.e.\ in general, $\Braket{\partial_\delta \Psi(t)|\partial_\delta \Psi(t)}\neq1$.}.  A detailed derivation of the QFI is provided in Appendix~\ref{appendix:QFI}. 

    \begin{figure*}
        \centering
        \includegraphics[width=1 \linewidth]{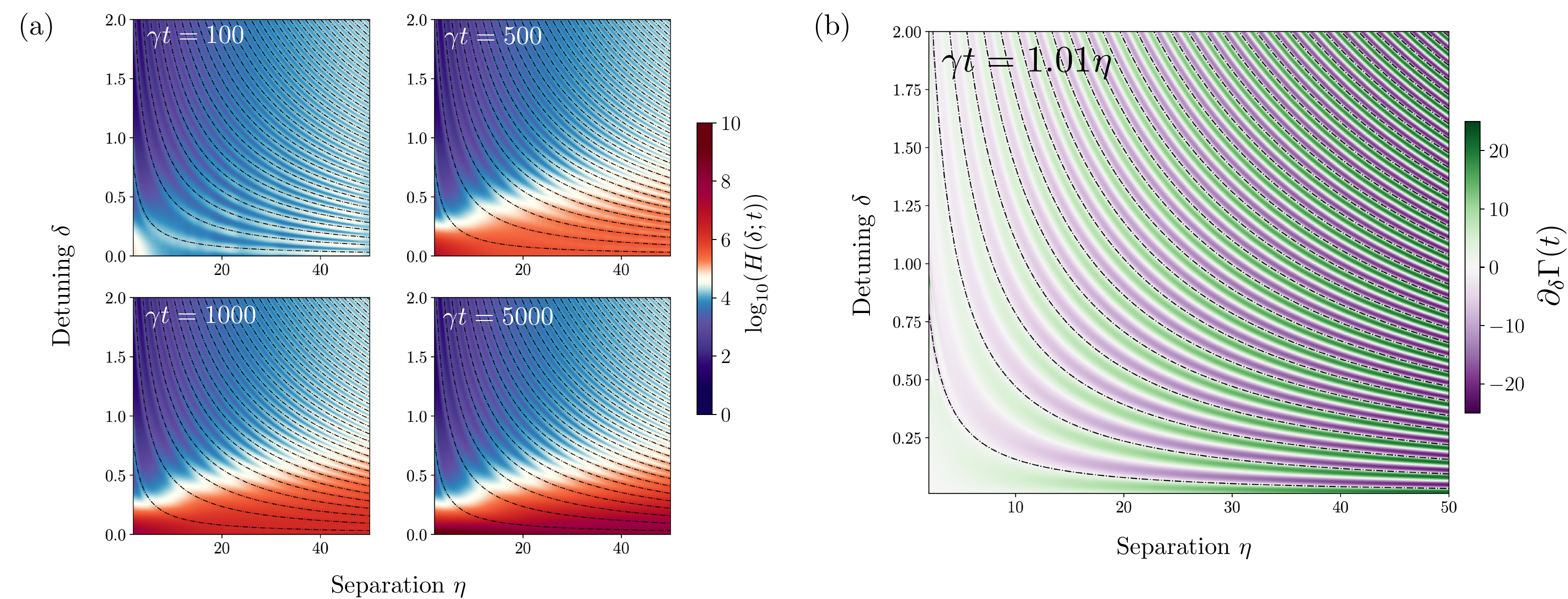}
        \caption{Quantum Fisher information ($H(\delta;t)$) and quasi-bound state decay rate derivative ($\partial_\delta \Gamma_{\rm qBIC}(t)$) for initial state $\ket{e,g}$ for varying atomic detuning ($\delta$) and separation ($\eta$). (a) QFI of the system with respect to $\delta$ evaluated at four times $\gamma t=100,500,1000,5000$; where steady state has been reached in the bottom right panel. (b) Derivative of the quasi-bound state excitation decay with respect to $\delta$, at a separation-dependent time $\gamma t\rightarrow\eta^+$, right after the field from emitter 1 reaches emitter 2. For each quantity, we note that the hyperbolae of constant propagation phase $\delta \eta =  (n+1/2)\pi$; $ n\in\mathbb{Z}^+$ (black dot-dashed lines) are local maxima.}
        \label{fig:qfi_bic}
    \end{figure*}  

    Fig.~\ref{fig:qfi_bic}\,(a) shows  the QFI $H(\delta;t)$  as a function of interatomic separations $(\eta)$ and detunings $(\delta)$ at various points in time (marked by the inset label), which is the central result of this work. We highlight below the key features of Fig.~\ref{fig:qfi_bic}\,(a), which can be broadly understood from  the build-up of field excitations inside the `cavity' formed by the two atoms, resulting in a longer interaction time over which the field interacts with the atoms before leaking out:
    
    \begin{itemize}
        \item {We note that the QFI generally increases with time, attaining a steady state value as $t\rightarrow\infty$. This can be attributed to the longer time of interaction between the emitters and the field that probes their relative detuning.      
        }
        \item{At a fixed time snapshot and  for a fixed atomic separation $\eta$, the QFI generally decreases with increasing the detuning $\delta$. The relative atomic detuning $ \delta $ affects the `cavity loss rate'. Thus for small $\delta$  the intra-cavity field decays slowly, allowing for longer interaction time between the field and the emitters, and consequently,  improved sensitivity of their relative detuning. }
        
        \item{For a fixed $\delta$, we identify two distinct trends for the QFI with varying atomic separation  $\eta$. First,  for small detunings (e.g., $\delta \lesssim 0.2$ region), we observe that the QFI generally decreases with the separation $ \eta$. This can be understood as follows: For small detunings and small distances, the two emitters form a near- perfect  bound state, which generally enhances the QFI.  However, with increasing the interatomic separation, some field leaks out of the `atomic cavity' before a quasi-bound state is formed. }
        \item{For larger detuning values $ (\delta\gtrsim1)$, e.g., when the emitter resonances are separated by a  linewidth or more, the multimode nature of the dynamics for large $ \eta$  starts to become significant. Particularly, we note that the QFI is maximum along hyperbolic curves of the $\eta$-$\delta$ plane. Specifically, these correspond to hyperbolae \eqn{\eta\delta= \bkt{n+\frac{1}{2}}\pi, \quad n\in \mathbb{Z}^+,
    \label{eq:etadelta2} }which is precisely the condition where resonance of emitter 1 aligns with the resonances of the `atomic cavity' (Eq.~\eqref{eq:deltaeta}). Such an alignment of atomic resonances with those of the `cavity' leads to a resonant build-up of field,  allowing the field to probe atoms for longer before it leaks. We will substantiate this observation  by analyzing the formation of quasi-bound states in the next section. 
    }

\begin{figure}[b]
    \centering
    \includegraphics[width=1\linewidth]{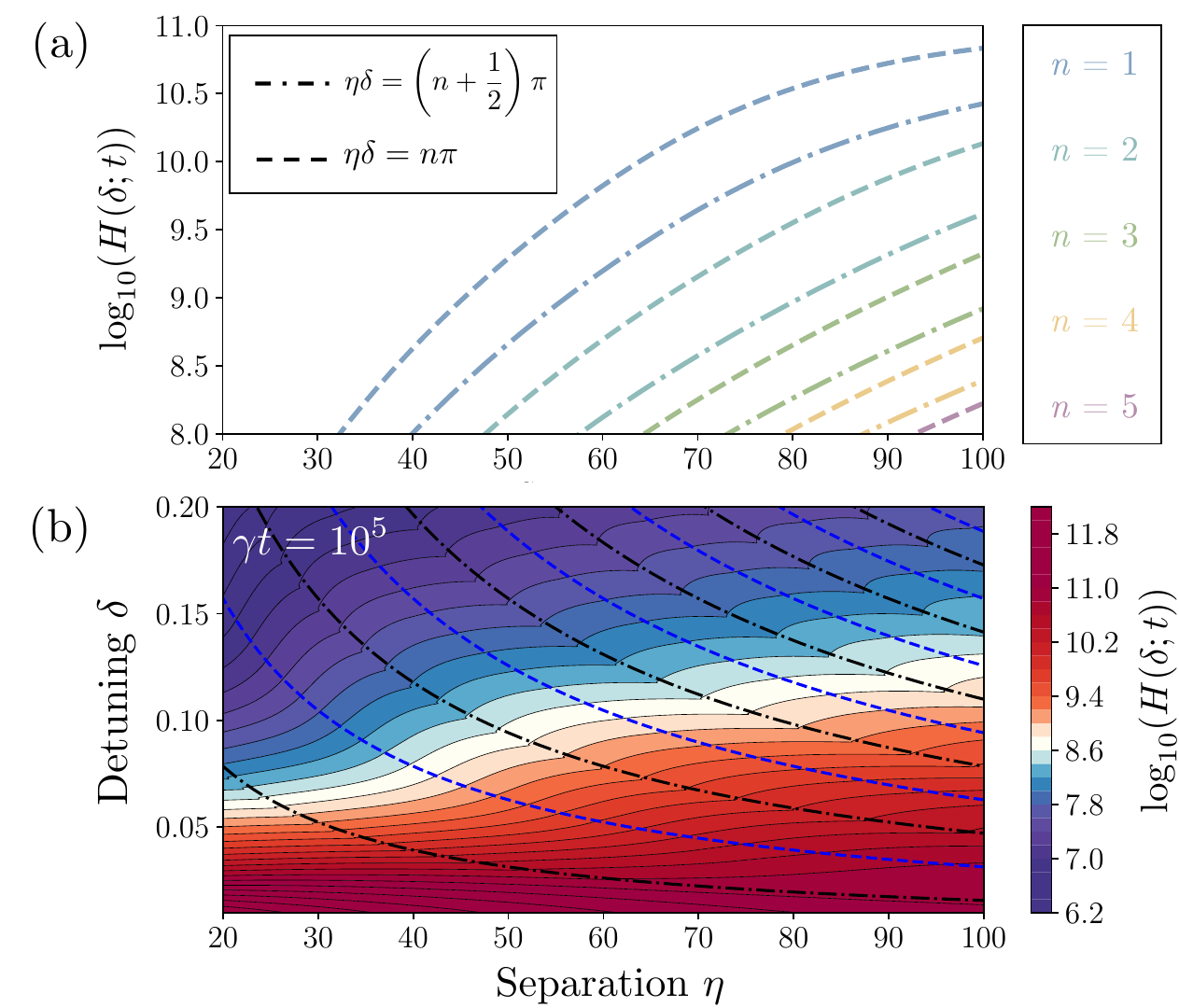}
    \caption{ Quantum Fisher information $H(\delta;t)$ for small detunings ($\delta\lesssim 0.2$)  at long times ($\gamma t = 10^5$).  (a) QFI along  $\eta\delta=(n+1/2)\pi$ (dot-dashed lines) and $\eta\delta=n\pi$ (dashed lines) hyperbolae. (b) QFI contour plot. For small detunings, the local maxima condition for QFI is at $\eta\delta=n\pi$, consistent with Eq.~\eqref{eq:deltaetasmall}.}
    \label{fig:qfi_small_delta}
\end{figure}
\item {
 Fig.~\ref{fig:qfi_small_delta} shows the QFI for small detunings $\delta\leq0.2$. We plot the QFI $H(\delta;t)$ as a function of the separation ($\eta$) along the hyperbolae $\eta\delta = \rm{\alpha}$ , where $\alpha = \cbkt{n\pi, \bkt{n + \frac{1}{2}\pi }}$, for $\gamma t =10^5$. We note that the QFI for $\alpha = n\pi$ (dashed lines) is consistently higher than that of the $\alpha = (n+1/2)\pi$ (dashed-dotted lines)  for the same $n\in \mathbb{N}$. This is also evident from the contour plot of $H(\delta;t)$ in Fig.~\ref{fig:qfi_small_delta}(b), where the local maxima shift to \eqn{\eta\delta=n\pi,}
  consistent with Eq.~\eqref{eq:deltaetasmall}, such that the detuned emitter frequency aligns with the effective cavity resonances. 
  The slower intra-cavity field decay in the small detuning regime is the primary reason the shift in the QFI maxima occurs at large times. 
} 
    \end{itemize}

    \subsection{Quasi-Bound State Formation}
    \label{sec:bic}

    As discussed in Sections~\ref{sec:emitter_dynamics} and~\ref{Sec:fielddyn}, the collective dynamics of two waveguide-coupled resonant  emitters separated by a long distance can result in a  delocalized atom-photon bound state,  arising from the destructive interference between the atoms and the fields propagating between them~\cite{Calajo2019, Sinha20a, KS2019}. 
     However, as noted previously, perfect BICs cannot emerge with finite detuning ($\delta\neq 0$) since the destructive interference between the fields emitted by each atom is no longer perfect, owing to their resonant frequency mismatch. While one can not have perfect bound states for two detuned emitters, the system still supports long-lived quasi bound states of the atoms and the field, which enable an enhancement in the resulting QFI. 
   
    To analyze such quasi-bound states, we consider  the total probability of the single excitation being in either of the two atoms, or in the waveguide field modes occupying the region between the two atoms $(x\in[-d/2,d/2])$~\cite{Xiong2022}:
    \begin{align}
        P_{q}(t)=\sum_{m=1,2}|c_m( t)|^2 + \int_{-d/2}^{d/2}\dd  x\, \sbkt{\abs{c_a( x,t)}^2 +\abs{c_b( x,t)}^2},
        \label{eq:prob_interatomic_main}
    \end{align}
    where 
    \begin{align}
        \begin{split}
            c_a( x,t) &= \frac{1}{2\pi}\int_0^\infty \dd \omega\, c_a( \omega, t) e^{-i\omega( t-x/v)},\\
        c_b( x, t) &= \frac{1}{2\pi}\int_0^\infty \dd \omega\, c_b( \omega, t) e^{-i\omega( t + x/v)}
        \end{split}
    \end{align}
    are the Fourier transforms of $c_a( \omega, t)$ and $c_b( \omega, t)$, respectively, switched from the frequency domain to spatial domain. 
    
    The time derivative of Eq.~\eqref{eq:prob_interatomic_main} gives us the excitation decay rate~\cite{Xiong2022} from the two-atoms + cavity region:
    \begin{align}
        \Gamma(t) & \equiv \frac{\dd P_q(t)}{\dd t} 
        \label{eq:loss_rate}
    \end{align}
    The values of $\eta$ and $\delta$ that minimize $\Gamma(t)$ correspond to a suppressed field leakage from the effective cavity formed by the two atoms, indicating the formation of a \emph{quasi-}bound state of the emitters and photons. In such a qBIC, the extended atom-field interaction time enables the field modes to interrogate the emitters for longer,  and consequently carry more information about the relative atomic detuning once they leak.
    
    We distinguish between two time regimes: before and after the initially emitted radiation of emitter 1 reaches emitter 2. Since we are interested in the late-time QFI, the decay rate for $t>d/v$ is given as

    \begin{align}
 &\Gamma\bkt{t>\frac{d}{v}} = \non\\
 &\sum_{m=1,2}\sbkt{\frac{\gamma}{8\pi}\cbkt{|c_m(t)|^2 -\left|c_m\left(t - \frac{d}{v}\right)\right|^2 }+\frac{\dd}{\dd t}\abs{c_m(t)}^2},
        \label{eq:gamma_a}
    \end{align}
    where the first line of the RHS corresponds to the total transfer of excitation probability from the emitters to the field in one round trip;  the second line represents the instantaneous rate  of excitation for the individual emitters. Taken together, these two contributions capture the decay rate of the total excitation in the emitters and the field that occupies the cavity region.  
    
    In Fig.~\ref{fig:qfi_bic}(b), we plot the gradient of the qBIC loss rate $\partial_\delta \Gamma(t\to (d/v)^+)$ for each point in the $\eta$-$\delta$ plane. We observe that the relation between $\eta$ and $\delta$ for which there exists a qBIC:
    \eqn{\eta\delta= \bkt{n+\frac{1}{2}}\pi}
    matches  the one for maximal QFI (Eq.~\eqref{eq:etadelta2}), as well as the condition for the atomic resonances to align with those of the `cavity' (Eq.~\eqref{eq:deltaeta}). Thus, as pointed earlier in Section~\ref{Sec:fielddyn}, the alignment between the emitter resonances with those of the cavity boosts the emission from the atoms into the field,  as well as supports the formation of a qBIC that enhances field buildup, prolonging atom-field interaction and thereby the QFI.
    The details of the calculation of~\eqref{eq:gamma_a}, and an analysis on the derivative of Eq.~\eqref{eq:gamma_a} with respect to $\delta$ are discussed in Appendix~\ref{Appendix:LossRate}. 

    \subsection{Benchmarking with Non-Interacting Atoms}
    \label{sec:bench}
    To provide a meaningful point of comparison for the metrological performance of our setup, we analyze a reference scenario consisting of two identical emitters, each coupled to an independent waveguide. In this configuration, the emitters do not interact and, consequently, their spatial separation does not influence the system dynamics. Hence, if the emitter is prepared in the $\ket{e,g}$ state (as we have assumed in our prior analysis), we can only ascertain the center wavelength of the excited atom. Thus, we seek to sense the center frequency of either atom, $\cbkt{\omega_1 = \omega_0 + \frac{\Delta \omega}{2},\,\omega_2 = \omega_0 - \frac{\Delta \omega}{2}}$, rather than their relative detuning $\delta$. The QFI with respect to\ $\omega_m$ of the joint system, labeled $H(\omega_m)$, lower bounds the minimum sensitivity of sensing the detuning as $\langle\Delta\delta^2\rangle\geq 2/H(\omega_m)$, where the additional factor of 2 arises from the need to perform two separate measurements (one for each atom).

    Considering the initial single excitation state, $\ket{\Psi(t=0)} = \sigma^{(m)}_+\ket{g,g,\{0\}_1,\{0\}_2}$, where
    $\ket{\{0\}_m}$ represents the waveguide modes coupled to the $m^{\text{th}}$ emitter, Wigner-Weisskopf theory gives us the joint state
    \begin{align}
        \begin{split}
            \ket{\Psi(t)} &= e^{-(\gamma/2 +i\omega_m)t} \sigma^{(m)}_+\ket{g,g}\otimes\ket{\{0\}_1,\{0\}_2} \\
        &\quad+ \ket{g,g}\otimes \int_0^t \dd\tau\ \xi(\tau) [a_k^\dagger(\tau)  +b_k^\dagger (\tau)]\,\ket{\{0\}_1,\{0\}_2},
        \end{split}
    \end{align}
    where $a_m^\dagger(\tau)$, $b_m^{\dagger}(\tau)$ are creation operations corresponding to left and right going temporal modes for the waveguide coupled to emitter $m$. The photon temporal wavepacket is given by
    $\xi(\tau)= \sqrt{\gamma/2}\,\exp({-(\gamma/2 +i\omega_m)\tau})$. 
    
    Since we assume perfect coupling to the waveguide modes, we may use the pure state QFI expression in Eq.~\eqref{eq:qfi_definition} evaluated with respect to\ $\omega_m$ (details in Appendix~\ref{app:qfi_baseline}), which yields
    \begin{align}
        H(\omega_m; t) = \frac{4}{\gamma^2}\left[1-(e^{-\gamma t} + 2\gamma t) e^{-\gamma t}\right].
    \end{align}
     In the limit $t\rightarrow\infty$, the QFI becomes $\lim_{t\rightarrow\infty} H(\omega_m) =4/\gamma^2$, which is a constant $H(\tilde{\omega}_m) =4$, when expressed in the normalized frequency units $\tilde{\omega}_m=\omega_m/\gamma$. The QFI of measuring the center frequency only depends on $\gamma$, and is independent of $\omega_m$, and consequently independent of $\Delta\omega$. Thus, the optimal sensitivity for measuring the atomic detuning as obtained from  the state of waveguide-coupled emitters surpasses that of independent non-interacting emitters.


     \section{Conclusions and outlook}
     \label{sec:conc}
     We  demonstrate that non-Markovian delay in waveguide QED can be a resource for distributed quantum sensing by investigating a minimal setup of two  waveguide-coupled atoms interacting with an external field. The external field introduces a detuning $ (\Delta \omega)$ between the two emitters, which we aim to sense. We show that the collective non-Markovian and multimode dynamics of the emitters in the presence of detuning exhibits  detuning-dependent revivals and suppression of  excitations (Fig.~\ref{fig:emitter_dynamics}). 
     The concomitant spectrum of the  field radiated by the emitters can be understood in terms of the effective `cavity' formed by the two emitters: when the effective resonances of such a cavity aligns with the emitter resonances, there is an enhanced emission of the radiated field and build-up of the field in the  region between the emitters for specific values of the emitter separations $(\eta = \gamma d/v)$ and their relative detunings $(\delta = \Delta \omega/\gamma)$, given by $\eta \delta \approx (n + \frac{1}{2}) \pi$ for $ \delta\gtrsim 1$, and $\eta \delta \approx n \pi$ for $ \delta\ll 1$  (Fig.~\ref{fig:spectrum}, and Eqs.~\eqref{eq:deltaeta} and \eqref{eq:deltaetasmall}).

     Our central result shows that the Quantum Fisher Information (QFI) for estimating the detuning parameter $ \delta$, which encodes the field gradient, is enhanced in the presence of non-Markovian delay (Fig.~\ref{fig:qfi_bic}(a)). Such an enhancement is linked to formation of atom-photon quasi-bound states: intuitively, the longer the photons spend in `cavity' formed by the atoms and `interrogate' the detuned emitters, the more information can be gained about the detuning parameter. We characterize such atom-photon quasi-bound states by the rate $ \Gamma(t)$ at which the excitation in the emitters + field in the cavity region decays. We identify the minimas in  $\Gamma$ as corresponding to formation of qBIC. The specific parameter values for which we observe peaks in the QFI align with the formation of quasi-bound states   (Figs.~\ref{fig:qfi_bic} and ~\ref{fig:qfi_small_delta}): $ \eta \delta = (n + \frac{1}2)\pi$ for $ \delta \gtrsim 1$, and $ \eta \delta =n\pi$ for $ \delta \ll 1$  ($n \in \mathbb{Z}^+$). Further, the formation of qBICs can be understood as arising from an alignment of the atomic resonances with the resonances of the cavity formed by the atoms. Thus the optimal sensing conditions correspond to delay and detuning parameter values which (1) support formation of quasi-bound states that facilitate a longer interrogation time, and (2) for long delays which bring in multiple modes of the environment to probe the collective dynamics of the emitters. Finally, we show that the QFI for estimating the relative atomic detuning as obtained from  the non-Markovian collective dynamics of waveguide-coupled emitters exceeds that of two independent emitters by several orders of magnitude.

The optimal (QFI-achieving) measurement will comprise a joint measurement of the waveguide modes and the atomic states.  The  information about the relative detuning of the atoms is encoded in the collective dyanmics of the emitters and, subsequently, in the excitations of the waveguide modes. The cavity-like resonance features (Fig.~\ref{fig:spectrum}) indicate that the optimal measurement will likely entail temporo-spectral processing of the waveguide excitation before a single photon measurement. Remarkably, recent works have shown that in the collective superradiant burst from an atomic ensemble, the temporal mode that carries optimal information, in fact, has a vanishing number of photons~\cite{Belliardo2026-ni}. We leave the derivation and constructive formulation of such an optimal measurement for future works.

Given the remarkable experimental progress over recent years, such long-distance waveguide QED regimes can be accessed across various platforms ranging from neutral atoms coupled to optical nanofibers~\cite{Pennetta2022, Lechner2023, Lee2025}, circuit QED and giant atoms~\cite{Mirhosseini2018, Kannan2020, Ferreira2021, Kannan2023, Storz2025, Chen2026, hernandez2026}, quantum dots coupled to nanophotonic waveguides~\cite{Kim2018, Tiranov2023}, waveguide lattices~\cite{Longhi20, Vicencio2025}, to matter waves in optical lattices~\cite{Krinner2018, Kim2025}. Notably,  Ref.~\cite{Kim2025} demonstrated that for ultracold atoms in a vertical one-dimensional optical lattice, emulating waveguide-coupled quantum emitters and photons, the propagation phase of the matter waves emulating photons depends on the gravity gradients. It would be pertinent to explore the role of non-Markovian collective dynamics exhibited in such a system as a resource for sensing gravitational field gradients.

The present framework can be generalized in various ways to further enhance the sensing capabilities of such waveguide QED systems. For example, extending the analysis to multi-emitter systems that act as additional probes~\cite{Dinc2019, Carmele2020,   Windt2025, capurso2025, Barahona-Pascual2026, ArranzRegidor2021}; dispersion engineering to create photonic environments that enable robust quasi-bound states~\cite{Hsu2016}; and driven emitters that allow for continuous probing of the system~\cite{DornerZoller, Sinha20b, ArranzRegidor2021}. 
     Overall, our results establish non-Markovian time-delayed feedback and multimode reservoirs  as a  resource for distributed quantum sensing, opening new avenues in quantum metrology with waveguide-coupled quantum emitters.

     \section{Acknowledgments}
     We acknowledge helpful discussions with H. Alaeian, A. Clerk,  C. N. Gagatsos, D. Schneble, P. Solano,  R. Trivedi and H. E. T\"ureci. P.D. acknowledges support from Engineering Research Center for Quantum Networks (CQN), awarded by the NSF and DoE under cooperative agreement number 1941583. K.S. acknowledges support from the National Science Foundation under Grant No. PHY-2418249,  the  Air Force Office of Scientific Research under Award No. FA9550-25-1-0333, the John Templeton Foundation under Award No. 63626, and the U.S. Department of Energy, Office of Science under Grant No. DESC0026059. This work was supported  by grant NSF PHY-2309135 to the Kavli Institute for Theoretical Physics (KITP).

     Items and technical data in this article have been reviewed and determined not to be subject to the International Traffic in Arms Regulations (ITAR).

	\onecolumngrid    

    \appendix
    \section{Non-Markovian dynamics of waveguide-coupled detuned emitters}
\label{app:emitter_dyn}
Given the total Hamiltonian (Eqs.~\eqref{eq:Hamiltonian_emitter}--\eqref{eq:Hamiltonian_interaction}), we derive the following coupled  equations of motion for the  atomic and field excitation amplitudes defined in Eq.~\eqref{eq:single_excitation_ansatz}:
\begin{subequations}
    \begin{align}
        \dot{c}_a(\omega,t)&=-i \left[c_1(t)g^*(\omega) e^{-i\omega x_1/v} e^{i(\omega-\omega_1)t} + c_2(t)g^*(\omega) e^{-i\omega x_2/v} e^{i(\omega-\omega_2)t}\right], \label{eq:field_left}\\
        \dot{c}_b(\omega,t)&=-i \left[c_1(t)g^*(\omega) e^{i\omega x_1/v} e^{i(\omega-\omega_1)t} + c_2(t)g^*(\omega) e^{i\omega x_2/v} e^{i(\omega-\omega_2)t}\right],\label{eq:field_right}\\
        \dot{c}_m(t)&=-i \int_{0}^\infty \dd\omega \,g(\omega) e^{-i(\omega-\omega_m)t} \left[c_a(\omega,t) e^{i\omega x_m/v} + c_b(\omega,t) e^{-i\omega x_m/v}\right] ;\quad m=\{1,2\} .\label{eq:emitters}
    \end{align}
\end{subequations}

Integrating Eqs.~\eqref{eq:field_left} and~\eqref{eq:field_right} yields 
\eqn{
 {c}_a(\omega,t)&=-ig^*(\omega) \int_0 ^t \dd \tau \left[c_1(\tau) e^{-i\omega x_1/v} e^{i(\omega-\omega_1)\tau} + c_2(\tau) e^{-i\omega x_2/v} e^{i(\omega-\omega_2)\tau}\right], \label{eq:field_left2}\\
        \dot{c}_b(\omega,t)&=-ig^*(\omega) \int_0 ^t \dd \tau  \left[c_1(\tau) e^{i\omega x_1/v} e^{i(\omega-\omega_1)\tau} + c_2(\tau)e^{i\omega x_2/v} e^{i(\omega-\omega_2)\tau}\right],\label{eq:field_right2}
}
Substituting the above into the emitters' equation of motion Eq.~\eqref{eq:emitters} gives the coupled dynamical equations for the two atoms Eq.~\eqref{eq:dc1dt} and \eqref{eq:dc2dt}.
The corresponding Laplace transformed excitation amplitudes $\mathscr{C}_{1,2}(s)$  are:
\begin{subequations}
    \begin{align}
    \label{eq:laplace_fin1_app}
        \gamma	\mathscr{C}_1(\tilde s)=\frac{c_1(0)(\tilde{s}-i\delta+1/2)- c_2(0) e^{i\phi_1} e^{-\eta\tilde{s}}/2}{(\tilde{s}+1/2)(\tilde{s}-i\delta+1/2)-( e^{i\phi_1}e^{-\eta\tilde{s}}/2)^2},\\
        \label{eq:laplace_fin2_app}
        \gamma	\mathscr{C}_2(\tilde s)=\frac{c_2(0)(\tilde{s}+i\delta+1/2)-c_1(0) e^{i\phi_2} e^{-\eta\tilde{s}}/2}{(\tilde{s}+1/2)(\tilde{s}+i\delta+1/2)- ( e^{i\phi_2} e^{-\eta\tilde{s}}/2)^2}
        ,
    \end{align}
\end{subequations}
which yield Eqs.~\eqref{eq:laplacec1} and \eqref{eq:laplacec2} in the main text for the given initial conditions $\bkt{c_1 (0) = 1,c_2 (0) = 0}$.
We now detail the solutions to the coupled atomic dynamics, as obtained from taking the inverse Laplace transform of the above Laplace coefficients using two approaches.

\subsection{Atomic dynamics: Series solution}
\label{appendix:series_solution}

We solve  the coupled dynamics of the atoms by expanding the denominators in the Laplace coefficients in terms of a series, which physically corresponds to an expansion in terms of the number of photon wavepacket oscillations. We rewrite the Laplace coefficient $\mathscr{C}_2(s)$ in Eq.\eqref{eq:laplace_fin2_app} as
\begin{align}
        \gamma	\mathscr{C}_2(s)
        &=\frac{-\beta e^{i\phi_2} e^{-\eta\tilde{s}}/2}{(\tilde{s}+1/2)(\tilde{s}+i\delta+1/2)}  \sbkt{1-\frac{\beta^2 ( e^{i\phi_2} e^{-\eta\tilde{s}}/2)^2} {(\tilde{s}+1/2)(\tilde{s}+i\delta+1/2)}}^{-1}\\
    &= - \sum_{n=0}^\infty \frac{\beta^{2n+1} e^{i(2n+1)\phi_2}}{2^{2n+1}} \frac{e^{-(2n+1)\eta \tilde{s} }}{{(\tilde{s}+1/2)^{n+1}(\tilde{s}+i\delta+1/2)^{n+1}}}
    \label{eq:c2_series_laplace}
\end{align}
To invert the above Laplace coefficient, we utilize the following Laplace transform relations~\cite{Abramowitz2012-iz}:
\begin{subequations}
    \begin{align}
     e^{-sa} F(s)&\xrightarrow{\mathcal{L}^{-1}}f(t-a) \Theta(t-a) \\
    \frac{\Gamma(n)}{(s+a)^n (s+b)^n} &\xrightarrow{\mathcal{L}^{-1}} \sqrt{\pi}\left(\frac{t}{a-b}\right)^{n-1/2} e^{-(a+b)t/2} I_{n-1/2} \left( \frac{a-b}{2}t \right);\quad (n>0).
\end{align}
\end{subequations}
where $I_n$ is the modified Bessel function of the first kind, and $\Gamma(n)=(n-1)!$ is the Gamma function.  Using these relations in Eq.~\eqref{eq:c2_series_laplace} yields the time dynamics of atom 2 is given by
\eqn{
c_2(t) & = -\sqrt{\pi} e^{-i\delta \gamma t/2} \sum_{n=0}^\infty \frac{\beta^{2n+1} e^{i(2n+1)(\phi_1+\phi_2)/2}}{2^{2n+1} n!}  e^{-(\gamma t-(2n+1)\eta)/2} \Theta[\gamma t-(2n+1)\eta]  \nonumber\\
    & \qquad \qquad \qquad  \times  \left(\frac{\gamma t-(2n+1)\eta}{i\delta}\right)^{n+1/2} I_{n+1/2} \left( \frac{i\delta(\gamma t-(2n+1)\eta)}{2} \right).
}

This yields Eq.~\eqref{eq:c2t} with the coefficient $ \Xi _n ^{(2)}(t)$: 
\eqn{
\Xi_n^{(2)}(t) \equiv& - \frac{\sqrt{\pi}}{2^{2 n + 1} n !} \bkt{\frac{\gamma t - (2 n + 1) \eta}{i \delta }}^{n + 1/2}I_{n+1/2} \left( \frac{i\delta(\gamma t-(2n+1)\eta)}{2} \right)
}

We now use Eq.~\eqref{eq:dc2dt} to determine the excitation amplitude for the first atom,  $c_1 (t)$, in terms of $ c_2 (t)$.
For $t<d/v$, we get $c_1(t)=c_1(0) e^{-\gamma t/2}$, i.e., atomic decay due to spontaneous emission. For $t>d/v$, we have instead:
\begin{align}
    c_1(t)= -\frac{1}{\beta} e^{-i\phi_1 } e^{i\delta (\gamma t+\eta)} c_2\left(\gamma t+\eta\right) -\frac{2}{\gamma\beta}  e^{-i\phi_1 } e^{i\delta (\gamma t+\eta)} \dot{c}_2\left(\gamma t+\eta\right).
\end{align}
We use the recursive definition of the derivative of modified Bessel functions from Ref.~\cite{Gradshteyn2014-dw} to simplify the derivative $\dot{c}_2(t+d/v)$:
\begin{align}
    2 \frac{\dd}{\dd z} I_{n}(z)= I_{n+1}(z) + I_{n-1}(z).
\end{align}
We thus obtain the  expression for $c_1(t)$:
\eqn{
c_1(t)
=& e^{-\gamma t/2} +   \sqrt{\pi}e^{i\delta \gamma t/2}\sum_{n=1}^\infty \frac{\beta^{2n} e^{in(\phi_1+ \phi_2)}}{2^{2n+1}n!} \Theta(\gamma t-2n\eta)e^{-(\gamma t-2n\eta)/2} \bkt{\frac{\gamma t-2n\eta}{i\delta}}^{n-1/2}\non\\
        & \times  \sbkt{ \cbkt{\frac{2n+1}{i\delta} - (\gamma t-2n\eta) } 
      I_{n+1/2} \bkt{ \frac{i\delta(\gamma t-2n\eta)}{2} }\right.\non\\
    &\quad \quad \left.+  \bkt{\frac{\gamma t-2n\eta}{2}}
      \cbkt{I_{n+3/2} \bkt{ \frac{i\delta(\gamma t-2n\eta)}{2}} + I_{n-1/2} \bkt{ \frac{i\delta(\gamma t-2n\eta)}{2} } }}.
}

This results in  Eq.~\eqref{eq:c1t} with the coefficient  $\Xi _n ^{(1)}(t) $ defined as:
\eqn{&\Xi_n ^{(1)}(t)\equiv \non\\
&\frac{\sqrt{\pi}}{2^{2n + 1}n!} \sbkt{ \bkt{\frac{2n+1}{i\delta} } \bkt{\frac{\gamma t-2n\eta}{i\delta}}^{n-1/2} 
      I_{n+1/2} \bkt{ \frac{i\delta(\gamma t-2n\eta)}{2} }- i\delta \bkt{\frac{\gamma t-2n\eta}{i\delta}}^{n+1/2} 
      I_{n+1/2} \bkt{ \frac{i\delta(\gamma t-2n\eta)}{2} }\right.\non\\
    &\quad \quad  \quad  \quad \left.+ \frac{i\delta}{2}  \bkt{\frac{\gamma t-2n\eta}{i\delta}}^{n+1/2} \cbkt{I_{n+3/2} \bkt{ \frac{i\delta(\gamma t-2n\eta)}{2}} + I_{n-1/2} \bkt{ \frac{i\delta(\gamma t-2n\eta)}{2} } }}.
    }

\subsection{Atomic dynamics: Multiexponential decay}
\label{App:multiexp}
As an alternate approach for solving the atomic dynamics, one can invert Eq.~\eqref{eq:laplace_fin1_app}--\eqref{eq:laplace_fin2_app} using Cauchy's residue theorem. 
In order to apply Cauchy's residue theorem, we need to numerically evaluate the poles of Eq.~\eqref{eq:laplace_fin1_app} and~\eqref{eq:laplace_fin2_app}, i.e., determine the roots of the denominators, 
\begin{subequations}
    \begin{align}
    f_1(\tilde{s}; \delta)=(\tilde{s}+1/2)(\tilde{s}-i\delta+1/2)-\beta^2( e^{i\phi_1}e^{-\eta\tilde{s}}/2)^2,\\
    f_2(\tilde{s}; \delta)=(\tilde{s}+1/2)(\tilde{s}+i\delta+1/2)-\beta^2( e^{i\phi_2}e^{-\eta\tilde{s}}/2)^2.
\end{align}
\end{subequations}
We note that for $\delta=0$, $f_m(\tilde{s};\delta)=0$ resemble the transcendental equation $w e^w=z$. Hence, the solutions to $f_m(\tilde{s};0) =0 $ are given by the  Lambert-W function solutions,
\eqn{
\tilde{s}_{\pm,k}^{(m)}=\frac{1}{\eta}W_k\bkt{\pm\frac{\eta \beta}{2}\exp(i\phi_{j}+\eta/2)} -\frac{1}{2} \equiv \bar{s}_{\sigma,k}^{(m)},
}
where the superscript $m=1,2$ is the atomic index, while the subscripts $\sigma\in\{+,-\}$ and $k\in\mathbb{Z}$ denote the branch of the Lambert-W solution. 
For the sake of brevity, we adopt a joint subscript notation $\boldsymbol{j} \equiv (\sigma,j)$, where $\sigma\in\{+,-\}$ and $j\in \mathbb{Z}$ for the residues and poles in the rest of the appendices as well as the main tex.

However, for $\delta\neq0$, the equations no longer have the standard transcendental form. The general solutions are expected to be perturbations (proportional to $\delta$) of $\bar{s}_{\boldsymbol{k}}^{(m)}$. 

Given a solution $\bar{s}_{\boldsymbol{k}}^{(m)} = {\tilde{s}}_0$ for the equation $f_m(\tilde{s};0)=0$, let us assume that the roots of $f_m(\tilde{s};\delta)$ are of the form $\tilde{s}_0+\alpha$, where $\alpha\in \mathbb{C}$ is $\delta$ and pole index $k$ dependent, with no analytic closed form. We use the properties for Lambert-W functions to derive the equivalent transcendental equation (for $\beta=1$) 
    \begin{align}
        \begin{split}
            f_m(\tilde{s}_0+\alpha;\delta)=0 & \Rightarrow (\tilde{s}_0+\alpha +1/2)(\tilde{s}_0+\alpha \pm  i\delta+1/2)-( e^{i\phi_j}e^{-\eta(\tilde{s}_0+\alpha)}/2)^2=0\\
        & \Rightarrow \alpha^2 + 2\alpha (\tilde{s}_0+1/2) +(1-e^{-2\eta\alpha})(\tilde{s}_0+1/2)^2 \pm i\delta (\tilde{s}_0+\alpha+1/2) =0,\label{eq:reframe_transcendental}
        \end{split}
    \end{align}
where we use the relation $[\exp(i\phi_j-\eta \tilde{s}_0)/2]^2= (\tilde{s}_0+1/2)^2$, to replace the exponential with a polynomial term. This is a re-framed transcendental equation in terms of the shift variable $\alpha$. 

The corresponding residues  ${R}_{\sigma,j}^{(m)}$ for the $j^\mr{th}$ pole associated with the inverse Laplace transform of $ \mathscr{C}^{(m)}$ is:
\begin{align}
         {R}_{\sigma,j}^{(m)} = \frac{c_m(0)(\tilde{s}_{\sigma,j}^{(m)}+(-1)^m i\delta+1/2)-c_{m'}(0) e^{i\phi_m} e^{-\eta\tilde{s}_{\sigma,j}^{(m)}}/2}{1+2\tilde{s}_{\sigma,j}^{(m)}+(-1)^m i\delta+\eta\left( e^{2i\phi_n-2\eta\tilde{s}_{\sigma,j}^{(m)}}/2\right)},
         \label{eq:residue_value}
    \end{align}


For the numerical evaluation of the poles, we solve Eq.~\eqref{eq:reframe_transcendental} to determine the $\delta$-dependent shifts to the Lambert-W based solutions. This technique is numerically more stable since $|\alpha|$ is typically a small number, where  we find $|\alpha/(\tilde{s}_0+1/2)|\sim \mathcal{O}(1) $. The scaling of $\alpha$ can be determined by evaluating roots of the approximate polynomial equation for Eq.~\ref{eq:reframe_transcendental}, obtained by a Taylor expansion of the $e^{-\eta\alpha}$ term.  Additionally, since we evaluate the shift for every branch of the Lambert-W function, we avoid inadvertently skipping poles that contribute to the overall system evolution. We use Muller's root-finding algorithm for our numerics. For all the numerical evaluations in this article we use relative error tolerance $\epsilon\leq 10^{-7}$, maximum iteration count $ N=10^4$ and choose initial shifts $x_{\{0,1,2\}} = \{-0.05,-0.01,-0.01i\}\times \delta/\eta$ to obtain stable numerical solutions.

\section{Atomic frequency response and spectrum of the emitted field }
\label{appendix:atomic_frequency_response}

The frequency response functions for each atom are defined as the Fourier transform of the atomic amplitudes $F_{m}(\tilde \omega)\equiv \int \dd t\, c_m(t) e^{i(\tilde \omega-\tilde \omega_m) \gamma t}$. These can be related to the Laplace coefficients in Eq.~\eqref{eq:laplacec1}
 and \eqref{eq:laplacec2} as follows:
 \begin{align}
\gamma F_1(\tilde\omega) = \gamma \mathscr{C}_1\bkt{- i \bkt{\tilde \omega - \tilde \omega_1}}=&\frac{c_1(0)(- i \bkt{\tilde \omega - \tilde \omega_1}-i\delta+1/2)- c_2(0) e^{i\phi_1} e^{i\eta\bkt{\tilde \omega - \tilde \omega_1}}/2}{\bkt{- i \bkt{\tilde \omega - \tilde \omega_1}+1/2}(- i \bkt{\tilde \omega - \tilde \omega_1}-i\delta+1/2)-( e^{i\phi_1}e^{i\eta \bkt{\tilde \omega - \tilde \omega_1}}/2)^2},\\
            \gamma F_2(\tilde\omega) = \gamma	\mathscr{C}_2\bkt{-i (\tilde \omega - \omega_2)}=&\frac{c_2(0)\bkt{-i (\tilde \omega - \omega_2)+i\delta+1/2}-c_1(0) e^{i\phi_2} e^{i\eta (\tilde \omega - \omega_2)}/2}{(-i (\tilde \omega - \omega_2)+1/2)(-i (\tilde \omega - \omega_2)+i\delta+1/2)- ( e^{i\phi_2} e^{i\eta (\tilde \omega - \omega_2)}/2)^2}.
            \label{eq:f1_f2_appendix}
\end{align}
For the given initial state $ (c_1(0) = 1,\, c_2(0) = 0)$, the above expressions simplify to Eqs.~\eqref{eq:f1} and \eqref{eq:f2}. Substituting $ \bar \omega \equiv \tilde \omega - \tilde\omega_0 $, we get:
    \eqn{
\gamma F_1(\bar\omega ) &= \frac{-i\bkt{\bar{\omega}+\delta/2}+1/2}{D(\bar\omega)};\label{eq:f1bar} \\
\gamma F_2(\bar\omega) &= -\frac{ e^{i\eta \bar\omega}e^{i\eta \delta/2}/2}{D(\bar\omega)},
\label{eq:f2bar}
}
where the denominator $ D(\bar \omega)$ is given by
\eqn{
D\bkt{\bar \omega } = \sbkt{-i\bar \omega + \frac{1} {2}}^2+\frac{\delta^2}{4} - \frac{e^{2i\eta \bar\omega} e^{i\eta \delta}}{4}.
}

\begin{figure}[t]
    \centering
    \includegraphics[width=\linewidth]{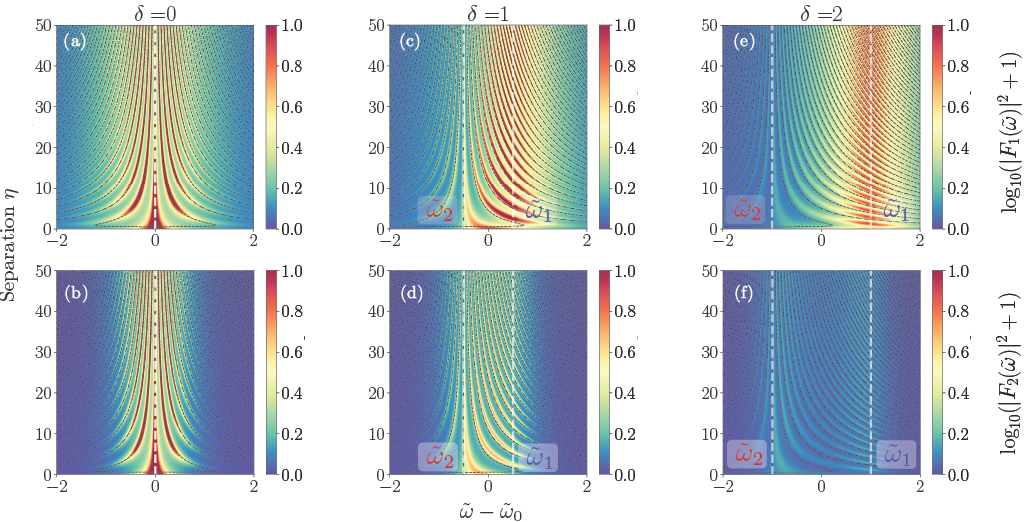}
    \caption{Atomic spectral response functions $ F_{1,2} (\tilde \omega)$  as a function of frequency $ \tilde \omega - \tilde \omega_0 $ and interatomic separation $ \eta$ for different values of detunings $ \delta$: $ \delta = 0 $ for (a) and (b); $\delta = 1 $ for (c) and (d); and $\delta = 2$ for (e) and (f).  The first row (a,c,e) shows the response functions for atom 1 ($ F_{1} (\tilde \omega)$), while the second row shows the response functions for atom 2 ($ F_{2} (\tilde \omega)$). The vertical white dashed lines indicate the frequencies of the two emitters. The black dash-dotted and dashed curves in Fig.~\ref{Fig:F1F2} correspond to the frequencies $ \tilde \omega_{n\delta} ^+ - \tilde \omega_0 $ and $ \tilde \omega_{n\delta} ^- - \tilde \omega_0 $, given by Eq.~\eqref{eq:baromega_pm_n}, respectively. Note that we have restricted the range of $z$-axis to keep consistent colorbar values for all plots for qualitative illustration. The $z$ ranges for the plots go above and below the set limits.}
    \label{Fig:F1F2}
\end{figure}

Fig.~\ref{Fig:F1F2} shows the atomic response functions for different detuning values. We note that the frequency response of emitter 1 ($F_1 \bkt{\tilde \omega}$) is peaked at its resonance $ \tilde\omega_1$, while that for emitter 2 ($F_2 \bkt{\tilde \omega}$) exhibits peaks at both $ \tilde \omega_1 $ and  $ \tilde \omega_2$. 

\subsection{Field spectrum }
The emitted field spectrum $ G(\tilde \omega)$, in the steady state, is given by 
    
    \eqn{G\bkt{\tilde \omega}\equiv  |c_{a}(\tilde \omega,\infty)|^2 + |c_{b}(\tilde \omega,\infty)|^2 =  \frac{\gamma}{\pi} \bkt{\abs{F_1 (\tilde\omega)}^2 + \abs{F_2 (\tilde\omega)}^2+ 2 \cos \bkt{\tilde \omega\eta}\re\sbkt{F_1 ^\ast \bkt{\tilde \omega}F_2 \bkt{\tilde \omega}} }.
    \label{eq:gw}
    }
Shifting the frequency axis to be centered at $ \tilde \omega_0 $, the explicit expression for the spectrum becomes:

\eqn{
G \bkt{\bar \omega} =  \frac{\gamma}{\pi} \sbkt{\abs{F_1 (\bar\omega )}^2 + \abs{F_2 (\bar\omega )}^2+ 2 \cos \bkt{(\bar \omega + \tilde \omega_0 )\eta}\re\sbkt{F_1 ^\ast \bkt{\bar \omega }F_2 \bkt{\bar \omega}} }.
}
Using  the condition 
\eqn{ \tilde \omega_0 \eta = \tilde \omega _2 \eta + \frac{\delta \eta } {2} = \phi_2  + \frac{\delta \eta } {2} = 2 p \pi + \frac{\delta \eta } {2} }
assuming the propagation phase $ \phi_2 = 2 p \pi$ (Eq.~\eqref{eq:phi2}), we get: 
\eqn{
G \bkt{\bar \omega} =  \frac{\gamma}{\pi} \sbkt{\abs{F_1 (\bar\omega )}^2 + \abs{F_2 (\bar\omega )}^2+ 2 \cos \bkt{\bar \omega\eta  + \frac{\delta \eta}{2}}\re\sbkt{F_1 ^\ast \bkt{\bar \omega }F_2 \bkt{\bar \omega}} }.
}

\subsection{Spectrum peaks in presence of detuning}
\label{appendix:eta_delta_condition}
Equating the approximate peaks of the spectrum obtained via Eq.~\eqref{eq:baromega_pm_n} to the resonance frequency of emitter 1, we get the following transcendental relation between the detuning $ \delta $ and the atomic separation $ \eta_n $
\begin{align}
    & \tilde \omega_{n \delta} ^\pm = \tilde \omega_1 \\
\implies & \tilde \omega_0 - \frac{\delta}{2}  +    \frac{1}{\eta_n}{\rm Im}\left[W_{n}\left(\pm\frac{\eta_n}{2}e^{\eta_n/2}\right)\right]= \tilde \omega_0 + \frac{\delta}{2}\\
    \implies & {\rm Im}\left[W_{n}\left(\pm\frac{\eta_n}{2}e^{\eta_n/2}\right)\right]=\delta \eta_n.
    \label{eq:deltaeta1}
\end{align}
In order to simplify the LHS of the above relation,  let us define 
\eqn{z _n^\pm  \equiv \pm \frac{\eta_n}{2} e^{\eta_n/2}.
}
For large atomic separations ($ \eta_n\gtrsim 1$), $ \abs {z_n^\pm}$ is also large such that  the Lambert-W function admits the asymptotic expansion~\cite{Corless1996}:
\begin{align}
    W_{n} (z_n^\pm)\approx\ln(z_n^\pm)+2in\pi-\ln\sbkt{\ln(z_n^\pm)+2in\pi}.
    \label{eq:lambert_asymptotic}
\end{align}

The imaginary part of the above simplifies to:
\eqn{
\im \sbkt{W_n \bkt{z_n^+ }} &\approx \im \sbkt{ \frac{\eta_n}{2} + \ln \bkt{\frac{\eta_n}{2}} + 2i n \pi - \ln \sbkt{\frac{\eta_n}{2} + \ln \bkt{\frac{\eta_n}{2}} + 2i n \pi }}\\
\implies\im \sbkt{W_n \bkt{z_n^+ }}& \approx  2 n \pi - \tan ^{-1 }\bkt{ \frac{2 n \pi }{\eta_n/2 + \ln (\eta_n/2)}},
\label{eq:wn+}
}
where we have used the simplification   $ \im [\ln (x + i y)] =\im \sbkt{\ln \bkt{\sqrt{x^2 + y^2} e^{i \tan^{-1} \bkt{y/x}}}} = \tan^{-1}(y/x)  $ in the second step.

Similarly for $ \im \sbkt{W_n \bkt{z_- }}$ we have:
\eqn{
\im \sbkt{W_n \bkt{z_n^- }} &\approx \im \sbkt{ \frac{\eta_n}{2} + \ln \bkt{\frac{\eta_n}{2}} + i (2n + 1) \pi - \ln \sbkt{\frac{\eta_n}{2} + \ln \bkt{\frac{\eta_n}{2}} + i (2n+ 1) \pi }}\\
\implies\im \sbkt{W_n \bkt{z_n^- }} &\approx  (2 n+1) \pi - \tan ^{-1 }\bkt{ \frac{(2 n+1) \pi }{\eta_n/2 + \ln (\eta_n/2)}}
\label{eq:wn-}
}

To further simplify the above expressions, we consider two parameter regimes:

\begin{enumerate}
    \item{Large detunings, $\delta> 1$: For $ \delta >1$, such that emitter detunings are larger than their respective  linewidths,  a large number ($n$) of branches of the Lambert-W function $ W_n(z_n^\pm) $ are required to describe the spectrum of the field close to the emitter resonances. This can be seen qualitatively from Fig.~\ref{fig:spectrum}. Thus for $ \delta \gtrsim  1$, the heuristic relation $ \delta \eta_n \sim n \pi $, implies that $ n \pi \gtrsim \eta_n $. In this regime,  $\tan^{-1} \bkt{ \frac{2 n \pi }{\eta_n/2 + \ln (\eta_n/2)}}\approx \pi/2$, such that Eqs.~\eqref{eq:wn+} and \eqref{eq:wn-} simplify to:  
    \eqn{ \im \sbkt{W_n \bkt{z_n^+ }} \approx 
    &\bkt{2 n  - \frac{1}{2}} \pi 
    \quad \implies \delta \eta_n \approx \bkt{2n - \frac{1}{2} } \pi 
    \label{eq:deleta+}\\
     \im \sbkt{W_n \bkt{z_n^- }} \approx &\bkt{2 n  + \frac{1}{2}} \pi \quad \implies \delta \eta_n \approx \bkt{2n + \frac{1}{2} } \pi ,
    \label{eq:deleta-}
    }
    where we have used the Eq.~\eqref{eq:deltaeta1}. Combining the above two conditions we get Eq.~\eqref{eq:deltaeta}.

    }
    \item{Small detunings, $\delta\ll 1$: For small emitter detunings, the corresponding  values of $ n $ for the Lambert-W function branches needed to describe the spectrum of the field near emitter resonances are small. Thus for $ \delta \ll  1$, the heuristic relation $ \delta \eta_n \sim n \pi $, implies that $ n \pi \ll \eta_n $. In this regime,  $\tan^{-1} \bkt{ \frac{2 n \pi }{\eta_n/2 + \ln (\eta_n/2)}}\approx 0$, such that Eqs.~\eqref{eq:wn+} and \eqref{eq:wn-} yield:  
    \eqn{ \im \sbkt{W_n \bkt{z_n^+ }} &\approx 
    2 n  \pi 
    \quad \implies \delta \eta_n \approx 2n   \pi 
    \label{eq:deleta+1}\\
     \im \sbkt{W_n \bkt{z_n^- }} &\approx \bkt{2 n  + 1} \pi \quad \implies \delta \eta_n \approx \bkt{2n + 1} \pi .
    \label{eq:deleta-1}
    }
    The above equations combined give us the condition: 
        \eqn{\eta_n \delta \approx n  \pi,
    \label{eq:deltaeta2}}
    yielding Eq.~\eqref{eq:deltaetasmall} in the main text.
    }
\end{enumerate}
For both cases the difference between adjacent peaks of the spectrum is  (from Eqs.~\eqref{eq:baromega_pm_n}, \eqref{eq:deleta+}, \eqref{eq:deleta-}, \eqref{eq:deleta+1} and \eqref{eq:deleta-1}):

\eqn{
\abs{\tilde  \omega_{n\delta} ^+ - \tilde \omega_{n\delta} ^-} = \frac{1}{\eta}\abs{\im \sbkt{W_n \bkt{\frac{\eta_n }{2} e^{\eta_n /2}} - W_n \bkt{-\frac{\eta_n }{2} e^{\eta_n /2}}}}
\approx \frac{\pi}{\eta}.
}
 This yields Eq.~\eqref{eq:spectrumFSR} in the main text.

\section{Quantum Fisher Information Evaluation}
\label{appendix:QFI}

We know that for a single parameter family of states $\rho_\theta$, the QFI can be directly calculated using the symmetric logarithmic derivative (SLD) $L_\theta$ by the relation
\begin{align}
	H(\theta) = \Tr[\rho_\theta L^2_\theta].
\end{align}
The SLD is implicitly defined by $\partial\rho_\theta/\partial \theta = \left(\rho_\theta L_\theta + L_\theta \rho_\theta\right)/2$. For pure states, i.e., states whose density matrix obeys $\rho_\theta^2 = \rho_\theta$~\cite{Fujiwara1995-kz}, we note that 
\begin{align}
    \frac{\partial \rho_\theta}{\partial \theta} =  \frac{\partial \rho_\theta^2}{\partial \theta} = \rho_\theta  \frac{\partial \rho_\theta}{\partial \theta} +  \frac{\partial \rho_\theta}{\partial \theta} \rho_\theta.
\end{align}
Thus, in this case $2{\partial \rho_\theta}/{\partial \theta}$ is the same as $L_\theta$. The quantum Fisher information is then 
\begin{align}
    H(\theta) =  \Tr[\rho_\theta L^2_\theta] &=  4 \Tr \rho_\theta \left(\frac{\partial \rho_\theta}{\partial \theta}\right)^2\\
    &=  4 \left[ \mathrm{Re} \Braket{\partial_\theta \psi_\theta|\partial_\theta \psi_\theta} + \Braket{\psi_\theta|\partial_\theta \psi_\theta}^2  \right].
\end{align}

For the present article, we seek to evaluate $H(\delta)$ to obtain bounds for sensing the atomic detuning. We define $\tilde \omega_1 =\tilde \omega_0 +\delta/2$ and $\tilde \omega_2 = \tilde \omega_0 -\delta/2$. Thus for $\ket{\partial_{\delta} \Psi(t) }$, we have
\begin{align}
    \frac{\partial\ket{\Psi_{\delta,\eta}(\tilde t)}}{\partial \delta} =\sum_{m={1,2}} \partial_{\delta} c_m(\tilde t)\hat{\sigma}_+^{(m)}\ket{g,g,\{0\}}+\int_0^\infty \dd\tilde \omega \, \left[\partial_{\delta}c_a(\tilde \omega,\tilde t)\hat{a}^\dagger(\tilde \omega)+\partial_{\delta}c_b(\tilde \omega,\tilde t)b^\dagger(\tilde \omega )\right] \ket{g,g,\{0\}}.
\end{align}

We now have to evaluate $\derdelta c_1(\tilde t), \derdelta c_2(\tilde t)$ based on Eq.~\ref{eq:residue_value}. We note that based on our numerical evaluation techniques for the poles in Appendix~\ref{app:emitter_dyn}, the poles $\tilde{s}_{\boldsymbol{k}}^{(m)}$ can be expressed as $\tilde{s}_{\boldsymbol{k}}^{(m)} = \bar{s}_{\boldsymbol{k}}^{(m)}+\alpha^{(m)}_{\boldsymbol{k}}(\delta)$ where $\bar{s}_{\boldsymbol{k}}^{(m)}$ is the $\delta=0$ solution and $\alpha^{(1)}_{\boldsymbol{k}}(\delta)$ is the $\delta$ dependent perturbation. As a reminder, we use the compact joint index $\boldsymbol{k} =(\sigma, k) \in \{+,-\} \times \mathbb{Z}$; where we truncate $|k|\leq 250$ for numerical evaluations.

We may take a partial derivative of the transcendental equation, Eq.~\eqref{eq:reframe_transcendental} to get
\begin{align}
    &2\alpha^{(1)}_{\boldsymbol{k}}(\delta) \derdelta[\alpha^{(1)}_{\boldsymbol{k}}(\delta)] + 2 [\bar{s}^{(1)}_{\boldsymbol{k}}+1/2]\derdelta [\alpha^{(1)}_{\boldsymbol{k}}(\delta)] +2\eta \beta^2 \left(\bar{s}^{(1)}_{\boldsymbol{k}}+1/2\right)^2 e^{-2\eta \alpha^{(1)}_{\boldsymbol{k}}(\delta)} \derdelta[\alpha^{(1)}_{\boldsymbol{k}}(\delta)] \nonumber \\
    &\qquad \qquad - i\left(\bar{s}^{(1)}_{\boldsymbol{k}}+\alpha^{(1)}_{\boldsymbol{k}}(\delta) +1/2\right) - i\delta \derdelta[\alpha^{(1)}_{\boldsymbol{k}}(\delta)] =0.
\end{align}
From which we obtain
\begin{align}
\derdelta[\alpha^{(1)}_{\boldsymbol{k}}(\delta)] = \frac{ i\left(\bar{s}^{(1)}_{\boldsymbol{k}}+\alpha^{(1)}_{\boldsymbol{k}}(\delta) +1/2\right)}{ 2\alpha^{(1)}_{\boldsymbol{k}}(\delta)+ 2 [\bar{s}^{(1)}_{\boldsymbol{k}}+1/2] + 2\eta \beta^2 \left(\bar{s}^{(1)}_{\boldsymbol{k}}+1/2\right)^2 e^{-2\eta \alpha^{(1)}_{\boldsymbol{k}}(\delta)} - i\delta }.
\end{align}
Similarly,
\begin{align}
    &\derdelta[\alpha^{(2)}_{\boldsymbol{k}}(\delta)] = \frac{ -i\left(\bar{s}^{(2)}_{\boldsymbol{k}}+\alpha^{(2)}_{\boldsymbol{k}}(\delta) +1/2\right)}{ 2\alpha^{(2)}_{\boldsymbol{k}}(\delta)+ 2 [\bar{s}^{(2)}_{\boldsymbol{k}}+1/2] + 2\eta \beta^2 \left(\bar{s}^{(2)}_{\boldsymbol{k}}+1/2\right)^2 e^{-2\eta \alpha^{(2)}_{\boldsymbol{k}}(\delta)} + i\delta }.
\end{align}
Hence, we may write $\derdelta c_1(\tilde t)$ as 
\begin{align}
    \begin{split}
        \derdelta c_1(\tilde{t}) &= \sum_{\boldsymbol{k}} \left[\frac{\mathscr{A}_{1,\boldsymbol{k}} - \mathscr{B}_{1,\boldsymbol{k}}}{\left(1+2\tilde{s}_{\boldsymbol{k}}^{(1)}-i\delta+\beta^2\eta( e^{2i\phi_2-2\eta\tilde{s}_{\boldsymbol{k}}^{(1)}}/2) \right)^2} + R_{\boldsymbol{k}}^{(1)}  \derdelta[\alpha_{\boldsymbol{k}}^{(1)}(\delta)]\right] \exp(\tilde{s}_{\boldsymbol{k}}^{(1)}t) \\
    &\equiv \sum_{\boldsymbol{k}} \mathcal{R}_{\boldsymbol{k}}^{(1)} \exp(\tilde{s}_{\boldsymbol{k}}^{(1)}t)
    \end{split}
\end{align}
where $\mathscr{A}_{1,\boldsymbol{k}}$ and $\mathscr{B}_{1,\boldsymbol{k}}$ are 
\begin{subequations}
    \begin{align}
    \mathscr{A}_{1,\boldsymbol{k}}&= \left[c_1(0)\left(\derdelta[\alpha_{\boldsymbol{k}}^{(1)}(\delta)]-i\right) -\beta c_2(0)  e^{i\phi_1-\eta\tilde{s}_{\boldsymbol{k}}^{(1)}}(i\eta/2 -\eta\derdelta[\alpha_{\boldsymbol{k}}^{(1)}(\delta)] )/2\right] \nonumber \\&\qquad\qquad\times \left(1+2\tilde{s}_{\boldsymbol{k}}^{(1)}-i\delta+\beta^2\eta( e^{2i\phi_2-2\eta\tilde{s}_{\boldsymbol{k}}^{(1)}}/2) \right) \\
    \mathscr{B}_{1,\boldsymbol{k}} &= \left(c_1(0)(\tilde{s}_{\boldsymbol{k}}^{(1)}-i\delta+1/2)-\beta c_2(0) e^{i\phi_1} e^{-\eta\tilde{s}_{\boldsymbol{k}}^{(1)}}/2\right) \nonumber \\ &\qquad\qquad\times \left[2\derdelta[\alpha_{\boldsymbol{k}}^{(1)}(\delta)] -i +\beta^2\eta ( e^{2i\phi_2-2\eta\tilde{s}_{\boldsymbol{k}}^{(1)}}/2)(-i\eta-2\eta \derdelta[\alpha_{\boldsymbol{k}}^{(1)}(\delta)])\right]
\end{align}
\end{subequations}
Similarly we have 
\begin{align}
    \begin{split}
         \derdelta c_2(\tilde{t}) &= \sum_{\boldsymbol{k}} \left[\frac{\mathscr{A}_{2,\boldsymbol{k}} - \mathscr{B}_{2,\boldsymbol{k}}}{\left(1+2\tilde{s}_{\boldsymbol{k}}^{(2)}+i\delta+\beta^2\eta( e^{2i\phi_1-2\eta\tilde{s}_{\boldsymbol{k}}^{(2)}}/2) \right)^2} + R_{\boldsymbol{k}}^{(2)}  \derdelta[\alpha_{\boldsymbol{k}}^{(2)}(\delta)]\right] \exp(\tilde{s}_{\boldsymbol{k}}^{(2)}t) \\
    &\equiv \sum_{\boldsymbol{k}} \mathcal{R}_{\boldsymbol{k}}^{(2)} \exp(\tilde{s}_{\boldsymbol{k}}^{(2)}t)
    \end{split}
\end{align}
where $\mathscr{A}_{2,\boldsymbol{k}}$ and $\mathscr{B}_{2,\boldsymbol{k}}$ are 
\begin{subequations}
    \begin{align}
    \mathscr{A}_{2,\boldsymbol{k}}&= \left[c_2(0)\left(\derdelta[\alpha_{\boldsymbol{k}}^{(2)}(\delta)]+i\right) -\beta c_1(0)  e^{i\phi_2-\eta\tilde{s}_{\boldsymbol{k}}^{(2)}}(-i\eta/2-\eta\derdelta[\alpha_{\boldsymbol{k}}^{(2)}(\delta)] )/2\right] \nonumber\\&\qquad\qquad\times \left(1+2\tilde{s}_{\boldsymbol{k}}^{(2)}+i\delta+\beta^2\eta( e^{2i\phi_1-2\eta\tilde{s}_{\boldsymbol{k}}^{(2)}}/2) \right) \\
    \mathscr{B}_{2,\boldsymbol{k}} &= \left(c_2(0)(\tilde{s}_{\boldsymbol{k}}^{(2)}+i\delta+1/2)-\beta c_1(0) e^{i\phi_2} e^{-\eta\tilde{s}_{\boldsymbol{k}}^{(2)}}/2\right) \nonumber\\&\qquad\qquad\times \left[2\derdelta[\alpha_{\boldsymbol{k}}^{(2)}(\delta)] +i +\beta^2\eta ( e^{2i\phi_1-2\eta\tilde{s}_{\boldsymbol{k}}^{(2)}}/2)(i\eta-2\eta \derdelta[\alpha_{\boldsymbol{k}}^{(1)}(\delta)])\right]
\end{align}
\end{subequations}

The derivatives of the field amplitudes can be calculated from
{\small
\begin{subequations}
    \begin{align}
    \derdelta c_a(\omega,t) = -i\sum_{\boldsymbol{k}} \int_{0}^t \, \dd \tau \, g^{*} (\omega) \left[ \left(\mathcal{R}_{\boldsymbol{k}}^{(1)}-i\frac{\tau}{2} R_{\boldsymbol{k}}^{(1)}\right) e^{\tilde{s}_{\boldsymbol{k}}^{(1)}\tau} e^{-i\omega x_1/v} e^{i(\omega-\omega_1)\tau} + \left(\mathcal{R}_{\boldsymbol{k}}^{(2)}+i\frac{\tau}{2} R_{\boldsymbol{k}}^{(2)}\right) e^{\tilde{s}_{\boldsymbol{k}}^{(2)}\tau} e^{-i\omega x_2/v} e^{i(\omega-\omega_2)\tau}\right]; \\
    \derdelta c_b(\omega,t) = -i\sum_{\boldsymbol{k}} \int_{0}^t \, \dd \tau \, g^{*} (\omega) \left[ \left(\mathcal{R}_{\boldsymbol{k}}^{(1)}-i\frac{\tau}{2} R_{\boldsymbol{k}}^{(1)}\right) e^{\tilde{s}_{\boldsymbol{k}}^{(1)}\tau} e^{i\omega x_1/v} e^{i(\omega-\omega_1)\tau} + \left(\mathcal{R}_{\boldsymbol{k}}^{(2)}+i\frac{\tau}{2} R_{\boldsymbol{k}}^{(2)}\right) e^{\tilde{s}_{\boldsymbol{k}}^{(2)}\tau} e^{i\omega x_2/v} e^{i(\omega-\omega_2)\tau}\right].
\end{align}
\end{subequations}
}
Now, for the required overlap integrals, we need to evaluate
\begin{subequations}
    \begin{align}
    \Braket{\Psi_{\delta,\eta}(t)|\derdelta\Psi_{\delta,\eta}(t)}_{\delta} & = \sum_{m={1,2}} c^{*}_m(t) \derdelta c_m(t) + \int_0^\infty \dd \omega \, \left[c^{*}_a(\omega,t) \, \derdelta c_a(\omega,t)+c^{*}_b(\omega,t) \, \derdelta c_b(\omega,t)\right]\\
    \Braket{\derdelta\Psi_{\delta,\eta}(t)|\derdelta\Psi_{\delta,\eta}(t)} & = \sum_{m={1,2}} \derdelta c^{*}_m(t) \derdelta c_m(t) + \int_0^\infty \dd \omega \, \left[\derdelta c^{*}_a(\omega,t) \cdot \derdelta c_a(\omega,t) + \derdelta c^{*}_b(\omega,t) \cdot \derdelta c_b(\omega,t)\right]
\end{align}
\end{subequations}
In the evaluation of these overlaps, we will encounter exponential integrals of the general form, $\int_{\tau_2}^{\tau_1} dt \, t^n \exp(-\alpha t)$ over a finite domain -- for the sake of brevity, we adopt the following shorthand notation in the following subsections,
\begin{align}
    \begin{split}
    &\mathcal{G}_1(u;\tau) = \frac{e^{u\tau}}{u}; \\
    &\mathcal{G}_2(u;\tau) = \frac{e^{u\tau}(u\tau-1)}{u^2} \\
    &\mathcal{G}_3(u;\tau) = \frac{e^{u\tau} {(u^2\tau^2 -2u\tau+2)}}{u^3}
    \end{split}
\end{align}
We will state the results of the $\int_{0}^{\infty} d\omega \, c^{*}_a(\omega,t) \derdelta c_a(\omega,t)$  and $\int_{0}^{\infty} d\omega \, \derdelta c^{*}_a(\omega,t) \derdelta c_a(\omega,t)$. The integrals for $c_b(\omega,t)$ can be obtained by setting $v\rightarrow-v$. 
We note that the integrands can be expressed in terms of the previously derived quantities as 
    \begin{align}
        &c^{*}_m(t) \derdelta c_m(t) =  \sum_{\boldsymbol{k}_1,\boldsymbol{k}_2} R_{\boldsymbol{k}_1}^{(m)*} \mathcal{R}_{\boldsymbol{k}_2}^{(m)}\exp\left(\left(\tilde{s}_{\boldsymbol{k}_1}^{(1)*}+\tilde{s}_{\boldsymbol{k}_2}^{(2)}\right)\tilde{t}\right)\\
        &\derdelta c^{*}_m(t) \derdelta c_m(t) = \sum_{\boldsymbol{k}_1,\boldsymbol{k}_2} \mathcal{R}_{\boldsymbol{k}_1}^{(m)*} \mathcal{R}_{\boldsymbol{k}_2}^{(m)}\exp\left(\left(\tilde{s}_{\boldsymbol{k}_1}^{(1)*}+\tilde{s}_{\boldsymbol{k}_2}^{(2)}\right)\tilde{t}\right)
    \end{align}
The evaluated integrals are expressed below. We adopt the joint subscript $\boldsymbol{j} \equiv (\sigma, j)$ where $\sigma\in\{+,-\}$ and $j\in\mathbb{Z}$, as defined in Appendix~\ref{app:emitter_dyn}. For practical evaluation, we finitely terminate our summation $|j|\leq j_{\rm cut}=250$.
 { 
 \begin{align}   
        \begin{split}
            \int_{0}^{\infty} \dd \omega \, c^{*}_a(\omega,t) \derdelta c_a(\omega,t) = & \frac{\gamma}{2} \sum_{\boldsymbol{k}_1,\boldsymbol{k}_2} \, R_{\boldsymbol{k}_1}^{(1)*} \mathcal{R}_{\boldsymbol{k}_2}^{(1)}\left[ \mathcal{G}_1\left(\tilde{s}_{\boldsymbol{k}_1}^{(1)*}\tau + \tilde{s}_{\boldsymbol{k}_2}^{(1)},\tau\right)\right]^{t}_{0} -\frac{i R_{\boldsymbol{k}_1}^{(1)*} {R}_{\boldsymbol{k}_2}^{(1)} }{2} \left[ \mathcal{G}_2\left(\tilde{s}_{\boldsymbol{k}_1}^{(1)*}\tau + \tilde{s}_{\boldsymbol{k}_2}^{(1)},\tau\right)\right]^{t}_{0}  \\
            & \qquad +\, R_{\boldsymbol{k}_1}^{(2)*} \mathcal{R}_{\boldsymbol{k}_2}^{(2)}\left[ \mathcal{G}_1\left(\tilde{s}_{\boldsymbol{k}_1}^{(2)}\tau + \tilde{s}_{\boldsymbol{k}_2}^{(2)},\tau\right)\right]^{t}_{0} +\frac{i R_{\boldsymbol{k}_1}^{(2)*} {R}_{\boldsymbol{k}_2}^{(2)} }{2} \left[ \mathcal{G}_2\left(\tilde{s}_{\boldsymbol{k}_1}^{(2)}\tau + \tilde{s}_{\boldsymbol{k}_2}^{(2)},\tau\right)\right]^{t}_{0} \\
            & \qquad +\, R_{\boldsymbol{k}_1}^{(1)*} \left[\mathcal{R}_{\boldsymbol{k}_2}^{(2)} +\frac{id}{2v} R_{\boldsymbol{k}_2}^{(2)}\right] e^{+\tilde{s}_{\boldsymbol{k}_2}^{(2)}d/v-i\omega_2d/v} \left[\mathcal{G}_1\left(\tilde{s}_{\boldsymbol{k}_1}^{(1)*}+ \tilde{s}_{\boldsymbol{k}_2}^{(2)}+ i\Delta  \omega,\tau\right)\right]^{t-d/v}_{0}\\
            & \qquad +\, \frac{i}{2} R_{\boldsymbol{k}_1}^{(1)*} R_{\boldsymbol{k}_2}^{(2)} e^{+\tilde{s}_{\boldsymbol{k}_2}^{(2)}d/v-i\omega_2d/v} \left[\mathcal{G}_2\left(\tilde{s}_{\boldsymbol{k}_1}^{(1)*}+ \tilde{s}_{\boldsymbol{k}_2}^{(2)}+ i\Delta  \omega,\tau\right)\right]^{t-d/v}_{0}\\
            & \qquad +\, R_{\boldsymbol{k}_1}^{(2)*} \left[\mathcal{R}_{\boldsymbol{k}_2}^{(1)} +\frac{id}{2v} R_{\boldsymbol{k}_2}^{(1)}\right] e^{-\tilde{s}_{\boldsymbol{k}_2}^{(1)}d/v+i\omega_1 d/v} \left[\mathcal{G}_1\left(\tilde{s}_{\boldsymbol{k}_1}^{(2)}+ \tilde{s}_{\boldsymbol{k}_2}^{(1)}- i\Delta  \omega,\tau\right)\right]^{t}_{d/v} \\
            & \qquad -\, \frac{i}{2} R_{\boldsymbol{k}_1}^{(2)*} R_{\boldsymbol{k}_2}^{(1)} e^{-\tilde{s}_{\boldsymbol{k}_2}^{(1)}d/v+i\omega_1 d/v} \left[\mathcal{G}_2\left(\tilde{s}_{\boldsymbol{k}_1}^{(2)}+ \tilde{s}_{\boldsymbol{k}_2}^{(1)}- i\Delta  \omega,\tau\right)\right]^{t}_{d/v}
        \end{split} 
    \end{align}
}
    Similarly, we evaluate $\int_{0}^{\infty} d\omega \, \derdelta c^{*}_a(\omega,t) \derdelta c_a(\omega,t)$ to obtain
     {
     \begin{align}       
        \begin{split}
            &\int_{0}^{\infty} \dd \omega \, \derdelta c^{*}_a(\omega,t) \derdelta  c_a(\omega,t) = \frac{\gamma}{2} \sum_{\boldsymbol{k}_1,\boldsymbol{k}_2}\mathcal{R}_{\boldsymbol{k}_1}^{(1)*}\mathcal{R}_{\boldsymbol{k}_2}^{(1)} \times \left[ \mathcal{G}_1\left(\tilde{s}_{\boldsymbol{k}_1}^{(1)*} + \tilde{s}_{\boldsymbol{k}_2}^{(1)};\; \tau\right)\right]_0^{t} \\
            &+\frac{i}{2} (R_{\boldsymbol{k}_1}^{(1)*}\mathcal{R}_{\boldsymbol{k}_2}^{(1)}-\mathcal R_{\boldsymbol{k}_1}^{(1)*}{R}_{\boldsymbol{k}_2}^{(1)}) \times \left[ \mathcal{G}_2\left(\tilde{s}_{\boldsymbol{k}_1}^{(1)*} + \tilde{s}_{\boldsymbol{k}_2}^{(1)};\; \tau\right)\right]_0^{t} + \frac{R_{\boldsymbol{k}_1}^{(1)*} R_{\boldsymbol{k}_2}^{(1)}}{4} \left[ \mathcal{G}_3\left(\tilde{s}_{\boldsymbol{k}_1}^{(1)*} + \tilde{s}_{\boldsymbol{k}_2}^{(1)};\; \tau\right)\right]_0^{t}\\
           &+\mathcal{R}_{\boldsymbol{k}_1}^{(2)*}\mathcal{R}_{\boldsymbol{k}_2}^{(2)} \times \left[ \mathcal{G}_1\left(\tilde{s}_{\boldsymbol{k}_1}^{(2)} + \tilde{s}_{\boldsymbol{k}_2}^{(2)};\; \tau\right)\right]_0^{t} \\
            &+\frac{i}{2} (R_{\boldsymbol{k}_1}^{(2)*}\mathcal{R}_{\boldsymbol{k}_2}^{(2)}-\mathcal R_{\boldsymbol{k}_1}^{(2)*}{R}_{\boldsymbol{k}_2}^{(2)}) \times \left[ \mathcal{G}_2\left(\tilde{s}_{\boldsymbol{k}_1}^{(2)} + \tilde{s}_{\boldsymbol{k}_2}^{(2)};\; \tau\right)\right]_0^{t} + \frac{R_{\boldsymbol{k}_1}^{(2)*} R_{\boldsymbol{k}_2}^{(2)}}{4} \left[ \mathcal{G}_3\left(\tilde{s}_{\boldsymbol{k}_1}^{(2)} + \tilde{s}_{\boldsymbol{k}_2}^{(2)};\; \tau\right)\right]_0^{t}\\
            & + \mathcal{R}_{\boldsymbol{k}_1}^{(1)*} \left[\mathcal{R}_{\boldsymbol{k}_2}^{(2)}+\frac{id}{2v} R_{\boldsymbol{k}_2}^{(2)}\right] e^{(\tilde{s}_{\boldsymbol{k}_2}^{(2)} -i \omega_2)d/v}  \times \left[ \mathcal{G}_1\left(\tilde{s}_{\boldsymbol{k}_1}^{(1)*} + \tilde{s}_{\boldsymbol{k}_2}^{(2)} + i\Delta\omega;\; \tau\right)\right]_{0}^{t-d/v} \\
            & + \frac{i}{2}(R_{\boldsymbol{k}_1}^{(1)*}\mathcal{R}_{\boldsymbol{k}_2}^{(2)} + \mathcal{R}_{\boldsymbol{k}_1}^{(1)*}{R}_{\boldsymbol{k}_2}^{(2)} ) e^{(\tilde{s}_{\boldsymbol{k}_2}^{(2)} -i \omega_2)d/v}  \times \left[ \mathcal{G}_2\left(\tilde{s}_{\boldsymbol{k}_1}^{(1)*} + \tilde{s}_{\boldsymbol{k}_2}^{(2)} + i\Delta\omega;\; \tau\right)\right]_{0}^{t-d/v} \\
            & -\frac{1}{4} R_{\boldsymbol{k}_1}^{(1)*} {R}_{\boldsymbol{k}_2}^{(2)} e^{(\tilde{s}_{\boldsymbol{k}_2}^{(2)} -i \omega_2)d/v}  \times \left[ \mathcal{G}_3\left(\tilde{s}_{\boldsymbol{k}_1}^{(1)*} + \tilde{s}_{\boldsymbol{k}_2}^{(2)} + i\Delta\omega;\; \tau\right)\right]_{0}^{t-d/v} \\
            & + \mathcal{R}_{\boldsymbol{k}_1}^{(2)*} \left[\mathcal{R}_{\boldsymbol{k}_2}^{(1)}+\frac{id}{2v} R_{\boldsymbol{k}_2}^{(1)}\right] e^{-(\tilde{s}_{\boldsymbol{k}_2}^{(1)} - i \omega_1)d/v}  \times \left[ \mathcal{G}_1\left(\tilde{s}_{\boldsymbol{k}_1}^{(2)} + \tilde{s}_{\boldsymbol{k}_2}^{(1)} - i\Delta\omega;\; \tau\right)\right]_{d/v}^{t} \\
            & - \frac{i}{2}(R_{\boldsymbol{k}_1}^{(2)*}\mathcal{R}_{\boldsymbol{k}_2}^{(1)} + \mathcal{R}_{\boldsymbol{k}_1}^{(2)*}{R}_{\boldsymbol{k}_2}^{(1)} )  e^{-(\tilde{s}_{\boldsymbol{k}_2}^{(1)} -i \omega_1)d/v}  \times \left[ \mathcal{G}_2\left(\tilde{s}_{\boldsymbol{k}_1}^{(2)} + \tilde{s}_{\boldsymbol{k}_2}^{(1)} - i\Delta\omega;\; \tau\right)\right]_{d/v}^{t} \\
            & - \frac{1}{4} R_{\boldsymbol{k}_1}^{(2)*} {R}_{\boldsymbol{k}_2}^{(1)}
            e^{-(\tilde{s}_{\boldsymbol{k}_2}^{(1)} -i \omega_1)d/v}  \times \left[ \mathcal{G}_3\left(\tilde{s}_{\boldsymbol{k}_1}^{(2)} + \tilde{s}_{\boldsymbol{k}_2}^{(1)} - i\Delta\omega;\; \tau\right)\right]_{d/v}^{t}
        \end{split}
    \end{align}
    }
    Note that in the normalized time variable $\tilde{t}$, all limits are replaced by their normalized versions, i.e.\ $d/v \equiv \eta$, and $t-d/v \equiv \tilde{t}-\eta$. Additionally, since the overlap integrals involve double time integrals, QFI expressed in normalized time differs from the standard time variable expression by a factor of $\gamma^2$, i.e. 
    \begin{align}
        H(\delta;\tilde{t}) = \gamma^2 H(\delta;{t}).
    \end{align}

\section{Atom-Photon Quasi-Bound State Formation }
\label{Appendix:LossRate}
In this Appendix, we expand on the steps to obtain the interatomic loss rate linked to the formation of a quasi-bound state in the continuum (qBIC) from the main text. The total probability of the single excitation being in either of the two atoms or in the field between the two atoms at any given time is
\begin{align}
    P_q( t)=\sum_{m=1,2}|c_m(t)|^2 + \int_{-d/2}^{d/2}\dd  x\, \big[|c_a( x, t)|^2 +|c_b( x,t)|^2\big],
    \label{eq:prob_interatomic}
\end{align}
where  $c_a( x, t) = \frac{1}{2\pi}\int_0^\infty \dd \omega\, c_a(\omega, t) e^{-i\omega\bkt{ t- x/v}}$ and   $c_b( x, t) = \frac{1}{2\pi}\int_0^\infty \dd \omega\, c_b( \omega, t) e^{-i\omega\bkt{ t+x/v}}$
are the inverse Fourier transforms of $c_a(\omega, t)$ and $c_b( \omega, t)$. We wish to compute the rate at which this excitation probability changes over time:
\begin{align}
    \frac{\dd P_q( t)}{\dd  t} 
    &= 2\re\left\{ \sum_{m=1,2} \dot{c}_m^*( t)c_m( t) +\int_{-d/2}^{d/2}\dd  x\, \left[\dot{c}_a^*(x, t)c_a(x, t)+\dot{c}_b^*( x, t)c_b(x, t)\right]\right\} \equiv \Gamma(t) .
\end{align}
The values of $\eta$ and $\delta$ that locally minimize $\Gamma( t)$ correspond to a slow excitation loss from the effective cavity formed by the two atoms, indicating the formation of a qBIC.
The transformation of the field amplitudes from frequency to position domain give us 
\begin{align}
    c_a( x, t)=-\frac{i}{2\pi}\int_0^\infty \dd  \omega~ g^*(\omega)~\int_0^{t}\dd\tau\,\left[ c_1 (\tau) e^{i\omega {d/2v}} e^{i(\omega-\omega_1)\tau}+ c_2 (\tau) e^{-i\omega{d/2v}} e^{i(\omega-\omega_2)\tau}\right] e^{-i \omega(  t-x/v)},\\
    c_b(x, t)=-\frac{i}{2\pi}\int_0^\infty \dd  \omega~ g^*(\omega)~\int_0^{ t}\dd\tau\,\left[ c_1 (\tau) e^{-i\omega{d/2v}} e^{i(\omega-\omega_1)\tau}+ c_2 (\tau) e^{i\omega{d/2v}} e^{i(\omega-\omega_2)\tau}\right] e^{-i \omega( t + x/v)}.
\end{align}
Again, assuming a flat-spectral density of the waveguide (i.e.\ $g({\omega}) = g({\omega}_0) $) and simplifying the integrals gives us the spatial domain field expressions,
\begin{subequations}
    \begin{align}
    c_a( x, t)=&-\frac{i g^*(\omega_0)}{2}\{c_1( t- (x-d/2)/v)e^{-i\omega_1(t- (x-d/2)/v)}[\Theta(t- (x-d/2)/v)-\Theta(- (x-d/2)/v)]\nonumber \\
    & +c_2( t- (x+d/2)/v)e^{-i\omega_2( t-(x+d/2)/v)}[\Theta( t- (x+d/2)/v)-\Theta(-(x+d/2)/v)]\},\label{eq:field_amp_spatial_a}\\
    c_b( x, t)=&-\frac{i g^*(\omega_0)}{2}\{c_1(t+ (x+d/2)/v)e^{-i\omega_1( t+ (x+d/2)/v)}[\Theta(t+ (x+d/2)/v)-\Theta( (x+d/2)/v)]\nonumber \\
    & +c_2( t+ (x-d/2)/v)e^{-i\omega_2( t+ (x-d/2)/v)}[\Theta( t+ (x-d/2)/v)-\Theta( (x-d/2)/v)]\}.
    \label{eq:field_amp_spatial_b}
\end{align}
\end{subequations}
In Eq. \eqref{eq:field_amp_spatial_a}, the $c_1(\cdot)$ term gives rise to fields emanated to the right from the atom at $x_1$ and the $c_2(\cdot)$ term gives rise to fields emanated to the right from the atom at $x_2$. Correspondingly, in Eq. \eqref{eq:field_amp_spatial_b}, these terms are fields emanating to the left from each atom. Consequently, we note that the second line of~\eqref{eq:field_amp_spatial_a} and first line of~\eqref{eq:field_amp_spatial_b} fall outside the region $x \in [-d/2,d/2]$, resulting in terms in the expansion of $|c_{\{a,b\}}(x, t)|^2$ that do not contribute to the integral of~\eqref{eq:prob_interatomic}: allowing us to simplify as
\begin{align}
    \int_{-{d/2}}^{{d/2}}\dd  x\, \sbkt{\abs{c_a\bkt{ x, t}}^2 +\abs{c_b\bkt{ x, t}}^2} =& \frac{\gamma}{8\pi} \int_{-{d/2}}^{{d/2}}\dd x\, \left\{ \abs{c_1( t- (x-d/2)/v)}^2 \sbkt{\Theta\bkt{ t-(x-d/2)/v}-\Theta\bkt{- (x-d/2)/v}}\right. \non \\ 
    &\left.+ \abs{c_2(t+ (x-d/2)/v)}^2 \sbkt{\Theta\bkt{t+ (x-d/2)/v}-\Theta\bkt{ (x-d/2)/v}}\right\} \equiv \mathscr{I}
\end{align}
The constraints imposed by the step function in the above expressions lead to two different results for times $0\le t\le d/v$ (radiation has not yet reached the second atom) and $t >d/v$ (radiation has reached the second atom and some is lost). Thus, the integral gets its limits rearranged as follows:
\begin{align}
    \mathscr{I} = \begin{cases}
        \frac{\gamma}{8\pi}  \left[\int_{-{d/2}}^{ vt-{d/2}}\dd  x\, |c_1( t- (x-d/2)/v)|^2 + \int_{{d/2}- vt}^{{d/2}}\dd  x\,|c_2( t+ (x-d/2)/v)|^2\right];  & t\in[0,d/v) \\[2ex]
        \frac{\gamma}{8\pi}\int_{-{d/2}}^{{d/2}}\dd  x  \left[\, |c_1( t- (x-d/2)/v)|^2 + \,|c_2(t+(x-d/2)/v)|^2\right]; & t>d/v
    \end{cases}
\end{align}
We simplify these integrals by substituting the atomic coefficients in their expanded forms
\begin{align}
        c_m({t}) 
        &\equiv\sum_{\boldsymbol{j}}^\infty {R}_{\boldsymbol{j}}^{(m)} e^{{s}_{\boldsymbol{j}}^{(m)} {t}},
\end{align}
The following simplified expressions are evaluated by taking the necessary definite spatial integrals
\begin{align}
    \label{Eq:FieldExcitationProbabilityBetweenTwoAtoms1}
    \mathscr{I}(0\leq t\leq d/v) 
    = \frac{\gamma}{8\pi} \sum_{\boldsymbol{j},\boldsymbol{k}} \left[-R_{\boldsymbol{j}}^{(1)}R_{\boldsymbol{k}}^{*(1)}/\cbkt{ s^{(1)}_{\boldsymbol{j}}+s^{(1)*}_{\boldsymbol{k}}}+R_{\boldsymbol{j}}^{(1)}R_{\boldsymbol{k}}^{*(1)}\exp\bkt{\cbkt{ s^{(1)}_{\boldsymbol{j}}+ s^{(1)*}_{\boldsymbol{k}}}t}/\cbkt{s^{(1)}_{\boldsymbol{j}}+ s^{(1)*}_{\boldsymbol{k}}} \non\right.\\
    \left.+ R_{\boldsymbol{j}}^{(2)}R_{\boldsymbol{k}}^{*(2)}\exp\bkt{\cbkt{ s^{(2)}_{\boldsymbol{j}}+ s^{(2)*}_{\boldsymbol{k}}}\tilde t}/\cbkt{ s^{(2)}_{\boldsymbol{j}}+ s^{(2)*}_{\boldsymbol{k}}}-R_{\boldsymbol{j}}^{(2)}R_{\boldsymbol{k}}^{*(2)}/\cbkt{ s^{(2)}_{\boldsymbol{j}}+ s^{(2)*}_{\boldsymbol{k}}} \right]
\end{align}
\begin{align}
    \label{Eq:FieldExcitationProbabilityBetweenTwoAtoms2}
    \mathscr{I}( t>d/v) 
    = \frac{\gamma}{8\pi} \sum_{\boldsymbol{j},\boldsymbol{k}} \left[-\cbkt{R_{\boldsymbol{j}}^{(1)}R_{\boldsymbol{k}}^{*(1)}\exp\bkt{\cbkt{ s^{(1)}_{\boldsymbol{j}}+s^{(1)*}_{\boldsymbol{k}}}\cbkt{ t-d/v}} -R_{\boldsymbol{j}}^{(1)}R_{\boldsymbol{k}}^{*(1)}\exp\bkt{\cbkt{s^{(1)}_{\boldsymbol{j}}+ s^{(1)*}_{\boldsymbol{k}}} t}}/\cbkt{ s^{(1)}_{\boldsymbol{j}}+s^{(1)*}_{\boldsymbol{k}}} \non\right.\\
    \left.+ \cbkt{R_{\boldsymbol{j}}^{(2)}R_{\boldsymbol{k}}^{*(2)}\exp\bkt{\cbkt{ s^{(2)}_{\boldsymbol{j}}+s^{(2)*}_{\boldsymbol{k}}}t}- R_{\boldsymbol{j}}^{(2)}R_{\boldsymbol{k}}^{*(2)}\exp\bkt{\cbkt{ s^{(2)}_{\boldsymbol{j}}+ s^{(2)*}_{\boldsymbol{k}}}\cbkt{ t-d/v}}}/\cbkt{ s^{(2)}_{\boldsymbol{j}}+ s^{(2)*}_{\boldsymbol{k}}} \right]
\end{align}
These equations \eqref{Eq:FieldExcitationProbabilityBetweenTwoAtoms1}, \eqref{Eq:FieldExcitationProbabilityBetweenTwoAtoms2} represent the probability of the single excitation being in the field modes between the two atoms. Now we can calculate the loss rate from the field modes by taking the time derivative:
\begin{align}
    \dot I_1(0\leq t\leq d/v)&=\frac{\gamma}{8\pi} \sum_{\boldsymbol{j},\boldsymbol{k}} \left[R_{\boldsymbol{j}}^{(1)}R_{\boldsymbol{k}}^{*(1)}\exp\bkt{\cbkt{ s^{(1)}_{\boldsymbol{j}}+ s^{(1)*}_{\boldsymbol{k}}} t}+ R_{\boldsymbol{j}}^{(2)}R_{\boldsymbol{k}}^{*(2)}\exp\bkt{\cbkt{ s^{(2)}_{\boldsymbol{j}}+ s^{(2)*}_{\boldsymbol{k}}} t} \right]\non\\
    &=\frac{\gamma}{8\pi}\sbkt{\abs{c_1(t)}^2+\abs{c_2(t)}^2},\\
    \dot I_2(t>d/v)&=\frac{\gamma}{8\pi} \sum_{\boldsymbol{j},\boldsymbol{k}} \left[-R_{\boldsymbol{j}}^{(1)}R_{\boldsymbol{k}}^{*(1)}\exp\bkt{\cbkt{ s^{(1)}_{\boldsymbol{j}}+ s^{(1)*}_{\boldsymbol{k}}}\cbkt{ t-d/v}} +R_{\boldsymbol{j}}^{(1)}R_{\boldsymbol{k}}^{*(1)}\exp\bkt{\cbkt{s^{(1)}_{\boldsymbol{j}}+ s^{(1)*}_{\boldsymbol{k}}} t} \non\right.\\
    &\left.+  R_{\boldsymbol{j}}^{(2)}R_{\boldsymbol{k}}^{*(2)}\exp\bkt{\cbkt{ s^{(2)}_{\boldsymbol{j}}+ s^{(2)*}_{\boldsymbol{k}}} t}- R_{\boldsymbol{j}}^{(2)}R_{\boldsymbol{k}}^{*(2)}\exp\bkt{\cbkt{ s^{(2)}_{\boldsymbol{j}}+ s^{(2)*}_{\boldsymbol{k}}}\cbkt{t-d/v}} \right]\non\\
    &=\frac{\gamma}{8\pi}\sbkt{-\abs{c_1( t-d/v) }^2+\abs{c_1(t)}^2 +\abs{c_2(t)}^2-\abs{c_2( t-d/v) }^2 }.
\end{align}
On the other hand, the loss rate of the atoms is given as $\Gamma_{\text {atoms} }=\partial_{ t} [|c_1( t)|^2+|c_2( t)|^2 ]=2\text{Re}\sbkt{\dot c_1(t)c_1^*( t)+\dot c_2( t)c_2^*( t)}$. Combining these results for field and atomic loss rates, we have
\begin{align}
    \Gamma( t)=
    \begin{cases}
    \displaystyle
        \sum_{m=1,2}\left[{2}\,\Re\!\left(\dot c_m^{*}(t)c_m(t)\right)+\frac{\gamma}{8\pi}\lvert c_m( t)\rvert^2\right],& 0 \le  t \le d/v, \\[2ex]
    \displaystyle\sum_{m=1,2}\left[{2}\,\Re\!\left(\dot c_m^{*}( t)c_m( t)\right)
    +\frac{\gamma}{8\pi}\!\left(\lvert c_m( t)\rvert^2-\lvert c_m( t-d/v)\rvert^2\right)\right], &  t > d/v , 
    \end{cases}
\end{align}
where the second line corresponds to Eq.~\eqref{eq:gamma_a}.
In order to determine qBIC conditions, we minimize the loss rate $\Gamma$ with respect to the detuning $\delta$. As we care about late-time dynamics, we only minimize the expression for $ t>d/v$. By substituting the expressions for the atomic dynamics, the extrema are found from the equation $\partial_\delta \Gamma( t>d/v)=0$.

\section{Quantum Fisher Information for Non-Interacting Emitters}
\label{app:qfi_baseline}
\begin{figure}[ht!]
    \centering
    \includegraphics[width=0.5\linewidth]{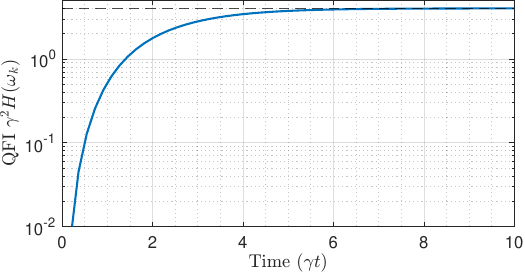}
    \caption{Plot of QFI $\gamma^2 H({\omega}_m)$ for non-interacting emitters as function of time.}
    \label{fig:qfi_baseline}
\end{figure}

We analyze a reference scenario consisting of two identical non-interacting emitters, each coupled to an independent waveguide.  Considering the initial state, $\ket{\Psi(t=0)} = \sigma^{(m)}_+\ket{g,g,\{0\}_1,\{0\}_2}$, where $\ket{\{0\}_m}$ represents the waveguide modes coupled to the $m^{\text{th}}$ emitter, Wigner-Weisskopf theory gives us the joint state
    \begin{align}
        \begin{split}
            \ket{\Psi(t)} = &e^{-(\gamma/2 +i\omega_m)t} \sigma^{(m)}_+\ket{g,g}\otimes\ket{\{0\}_1,\{0\}_2} + \ket{g,g}\otimes \int_0^t \dd\tau\, \xi(\tau) [a_m^\dagger(\tau)  +b_m^\dagger (\tau)]\,\ket{\{0\}_1,\{0\}_2}
        \end{split}
    \end{align}
    where the $a_m^\dagger(t')$, $b_m^{\dagger}(t')$ are creation operations corresponding to left and right going temporal modes for the waveguide coupled to emitter $m$. The photon temporal wavepacket is given by
    $\xi(t')= \sqrt{\frac{\gamma}{2}}e^{-(\gamma/2 +i\omega_m)t'}$, where $\omega_m$ is the center frequency and $\gamma$ is the overall atomic coupling rate to the waveguide. To evaluate the QFI, we calculate 
\begin{align}
    \ket{\partial_{\omega_m} \Psi(t)} = &-i t e^{-(\gamma/2 +i\omega_m)t} \sigma^{(m)}_+\ket{g,g}\otimes\ket{\{0\}_1,\{0\}_2} - \ket{g,g}\otimes \int_0^t \dd\tau\, i\tau \xi(\tau) [a_m^\dagger(\tau)  +b_m^\dagger (\tau)]\,\ket{\{0\}_1,\{0\}_2}
\end{align}
Evaluating the QFI $H(\omega_m) = 4(\mathrm{Re}\langle\partial_{\omega_m}\Psi(t)|\partial_{\omega_m}\Psi(t)\rangle + \langle\Psi(t)|\partial_{\omega_m}\Psi(t)\rangle^2)$ we get
\begin{align}
    \begin{split}
        \braket{\partial_{\omega_m}\Psi(t)} & = t^2 e^{-\gamma t} + 2\int_0^t \dd\tau\, \int_0^t \dd\tau'\, \tau \tau' \xi(\tau)\xi(\tau')\delta(\tau-\tau')= t^2 e^{-\gamma t}+ \gamma \int_0^t \dd\tau \, \tau^2 e^{-\gamma \tau} \\
    & = t^2 e^{-\gamma t} + \frac{2-e^{-\gamma t}(2+ 2\gamma t+\gamma^2 t^2)}{\gamma^2} = \frac{2(1-e^{-\gamma t}-\gamma t e^{-\gamma t})}{\gamma^2}
    \end{split}
\end{align}
\begin{align}
    \begin{split}
        \langle\Psi(t)|\partial_{\omega_m}\Psi(t)\rangle & = -i t e^{-\gamma t}  - 2i\int_0^t \dd\tau \, \int_0^t \dd\tau'\,  \tau' \xi(\tau)\xi(\tau')\delta(\tau-\tau')
    = -it e^{-\gamma t} -i\gamma \int_0^t \dd\tau \, \, \tau e^{-\gamma \tau} \\
    &=  -it e^{-\gamma t} -i \frac{1-e^{-\gamma t}(1+\gamma t)}{\gamma} =-i\frac{1-e^{-\gamma t}}{\gamma}
    \end{split}
\end{align}
which gives us  $ H(\omega_m) = 4\left(1-(e^{-\gamma t} +2\gamma t) e^{-\gamma t}\right)/\gamma^2$ (see plot in Fig.~\ref{fig:qfi_baseline}). It is important to note that the QFI is only determined by the coupling rate to the waveguide modes. In fact, if the waveguide modes are traced out, we have the state
\begin{align}
    \rho_{E_i} = e^{-\gamma t}\ket{e}\!\bra{e}_i +(1-e^{-\gamma t})\ket{g}\!\bra{g}_i
\end{align}
Since the state has no dependence on $\omega_i$, we have $\partial_{\omega_i}\rho_{E_i} = 0$, which consequently implies that the QFI is zero when the field modes are traced out.

We use the expression for the QFI of a parameter, $H(\theta) = H(f(\theta)) \times \left(\frac{d f(\theta)}{d\theta}\right)^2$, to obtain the normalized frequency units QFI as $\lim_{t\rightarrow\infty} H(\tilde{\omega}_m) = \gamma^2 \lim_{t\rightarrow\infty} H(\omega_m) =4$.

\twocolumngrid
        
     \bibliography{references}
    \balance

\end{document}